# Analysis of Transient Microscopy Data in Reciprocal Space

*Thomas J. Sheehan*[1] *and William A. Tisdale*[1]*

Author Addresses:

[1]Department of Chemical Engineering, Massachusetts Institute of Technology, Cambridge, Massachusetts 02139, United States

*email: tisdale@mit.edu

**Abstract**

Transient microscopy has evolved to include a family of well-established techniques for studying the spatiotemporal dynamics of excited states in a variety of materials systems. In these methods, a laser pulse is focused to locally excite a material, and the subsequent evolution of the excited species is visualized to extract transport parameters such as diffusivities. Typically, transport coefficients are extracted from experimental datasets using curve-fitting methods to analyze the change in the mean squared displacement (MSD) of the spatial profiles over time. However, this approach can lead to inaccurate or misleading diffusivity values under many common experimental conditions, and is only formally accurate for the case of a single diffusing species. In this work, we introduce a new approach to the analysis of transient microscopy data based on the evolution of profile shapes in reciprocal space. We use simulated datasets to quantitatively analyze the performance of our proposed methods and compare them to current state-of-the-art analysis techniques. We demonstrate that reciprocal-space-based analysis methods have the highest overall performance in terms of accuracy and precision for normal diffusive transport over a range of realistic initial conditions and signal-to-noise ratios. Furthermore, we highlight the limitations of using MSD alone to characterize anomalous transport in materials, and we show how analysis of spatial frequencies can yield a more comprehensive understanding of complex excited-state dynamics and their underlying physical mechanisms. Our results show that reciprocal-space-based analysis methods are powerful tools for characterizing and quantifying many different types of transport behaviors.

## 1. Introduction

Over the past decade, transient microscopy techniques have grown to become a powerful set of tools for measuring transport dynamics in a wide variety of materials, including both organic and inorganic semiconductors as well as structured metallic nanomaterials.[1–3] In these methods, a laser pulse is focused to locally excite a material, and the evolution of the excitation is then visualized to create a spatiotemporal map of energy flow in the material.[4,5] One reason for the widespread adoption of transient microscopy techniques is their flexibility. Time-resolved images have been collected using photoluminescence,[1,2] absorption,[6–11] and scattering-based setups,[12] allowing for measurements of the transport and dynamics of many different excited species, including excitons,[1,2,13–16] charge carriers,[17–20] heat or phonons,[3,12,21–23] polaritons,[24,25] and other hybrid quasiparticles.[26–28] Despite the wide variety of energy carriers and detection modalities, the quantitative analysis techniques used to extract diffusivities and other transport parameters from transient microscopy data remain relatively consistent.

The evolution of an excited species population is usually analyzed by fitting its spatial distribution at each moment in time to a shape having a time-dependent Gaussian component, such as a Gaussian or Voigt distribution.[29,30] The change in variance of the Gaussian component of the profile fits, which is equivalent to the mean squared displacement (MSD), is then used to quantitatively analyze the transport behavior of the excited species. In the case of diffusive transport along a given axis, the MSD increases linearly over time, with the diffusivity $D$ determined by the relationship:[4]

$$\mathrm{MSD} = 2\mathrm{Dt} \tag{1}$$

In its simplest form, this calculation procedure is only analytically valid for highly idealized behavior of excited species in the material. For example, it assumes that there is only one relevant species moving within the material, that the spatial profiles of the excited species match the specific functional form used to fit them (including both the shape of the initial profile and the Gaussian spreading over time), and that there are no non-linear recombination pathways which affect the profile shape. Practically, however, it is common for one or more of these assumptions to be violated. Given the prevalence of non-ideal diffusive behavior, more robust analysis techniques and experimental approaches are often needed to accurately determine transport parameters.

Many different approaches have been used to analyze complex behaviors in transient microscopy data. For example, to account for subtle differences between the shape of the initial excitation profile and the assumed shape used to determine MSDs, Hickey & Grumstrup proposed fitting spatial profiles at later times to a numerical convolution of the measured initial profile with a Gaussian spreading function.[29] Other works have accounted for non-Gaussian spreading of the excited species by using a numerical integration method to directly calculate the variances and MSDs of the spatial profile distributions at each point in time.[31,32] Rodríguez *et al.* have also proposed tracking the evolution of the excess kurtosis of profiles along with their MSDs to identify anomalous transport behavior.[32] In materials with multiple diffusing populations and non-linear MSD evolutions, it is common to calculate effective diffusivity values during different temporal regimes based on the local slope of the MSD curve.[33–36] Other works have attempted to model the dynamics of a composite system containing multiple species.[37–39] Finally, to account for non-linear interactions of excited species in the material, one can perform full 3D numerical simulations of the transport and recombination dynamics, and fit those predictions to experimental data.[17,18]

However, these existing analysis techniques often have limitations that can cause them to calculate inaccurate diffusivity values under certain conditions. For instance, in cases where spatial profiles are fit to specific distribution shapes to extract MSDs, even small deviations from the assumed profile shape can result in substantial errors in the calculated diffusivity values.[29] Most analysis techniques assume that the excited-state spatial profiles expand in a Gaussian manner, but this assumption generally does not hold when there is non-diffusive transport or when two or more types of excited species contribute to the microscopy data. As we show below, other MSD calculation techniques designed to be insensitive to non-idealities in profile shapes or non-Gaussian expansions, including the numerical convolution and numerical integration approaches, struggle with noisy data and can be challenging to implement on realistic datasets. Finally, we note that conventional analysis techniques do not extract all of the useful information present in transient microscopy data. The general framework of analyzing spatial profiles primarily through MSDs or other summary statistics often neglects the subtle but measurable ways that complex excited-state dynamics impact the overall profile shape. Clearly, there is room to learn even more from these rich spatiotemporal datasets.

In this paper, we introduce a framework that addresses many of the limitations of conventional data analysis techniques by analyzing transient microscopy data in reciprocal space. We outline a method of extracting MSDs and diffusivities based on changes in the relative intensities of different spatial frequencies over time. Using simulated datasets, we show that this method works for a variety of initial profile shapes and is less sensitive to noise than the numerical convolution and numerical integration approaches. Finally, we present an analysis framework in reciprocal space that extends beyond the ideal case of a single diffusing species, highlighting how different physical scenarios can be easily identified by their impacts on different spatial frequencies of the excited-state profiles. Our open-source code provides the community with a robust quantitative framework for analyzing transient microscopy data that can offer new insight into excited-state transport behaviors.

## 2. Analyzing the diffusion of a photoexcited species

### *2.1. Current State-of-the-Art in Transient Microscopy Data Analysis*

The dynamics of an excited species undergoing diffusion can be modeled using the following reaction-diffusion equation:

$$\frac{\partial n(\boldsymbol{x},t)}{\partial t} = D\nabla^2 n(\boldsymbol{x},t) - k_d\, n(\boldsymbol{x},t) - B\, n(\boldsymbol{x},t)^2 - C\, n(\boldsymbol{x},t)^3 \quad (2)$$

where $n$ is the excited-state distribution as a function of space $\boldsymbol{x}$ (a vector) and time $t$, $D$ is the diffusivity, assumed here to be constant and isotropic, $k_d$ is the first-order decay rate constant, and $B$ and $C$ are higher-order decay-rate constants.[18,40] For now, we focus on the low-fluence regime, which is frequently probed in transient microscopy experiments. Under these conditions, the impact of non-linear decay terms is assumed to be negligible, and the diffusion equation is reduced to:

$$\frac{\partial n(\boldsymbol{x},t)}{\partial t} = D\nabla^2 n(\boldsymbol{x},t) - k_d\, n(\boldsymbol{x},t) \quad (3)$$

This is a linear differential equation which is solved in 3D by:

$$n(\boldsymbol{x},t) = \frac{1}{\sqrt{(4\pi Dt)^3}} \exp[-k_d t]\, n(\boldsymbol{x},0) * \exp\left[-\frac{\|\boldsymbol{x}\|^2}{4Dt}\right] \quad (4)$$

Here, the $*$ symbol represents convolution over space. After normalization, this solution can be expressed as:

$$n(\boldsymbol{x},t) \propto n(\boldsymbol{x},0) * \exp\left[-\frac{\|\boldsymbol{x}\|^2}{4Dt}\right] = n(\boldsymbol{x},0) * \exp\left[-\frac{\|\boldsymbol{x}\|^2}{2\, MSD}\right] \quad (5)$$

In other words, the profile shape of an excited species as a function of time is given by the convolution of the original profile shape with a time-dependent Gaussian function. Along any given direction, the variance of the Gaussian component increases linearly with time according to the relationship $\Delta\sigma^2 = MSD = 2Dt$.

2.1.1. Choosing an initial profile shape

The most common approach to analyzing transient microscopy data is to fit the measured spatial profile at each point in time using the Gaussian expansion predicted by Eq. 5. Each spatial slice in time then yields one value of the mean-squared displacement of the population, or the MSD (see Table 1). In what we call the "Gaussian fitting method", the initial profile at $t = 0$ is assumed to have a Gaussian shape. Conveniently, subsequent profiles can also be fit to a Gaussian shape since the convolution of two Gaussians is another Gaussian. The MSDs can then be determined by the change in the variance of the Gaussian fits.[1]

In realistic scenarios using focused laser beams, the initial spatial profile of the photoexcited population will deviate from the simple Gaussian approximation. A popular alternative for one-dimensional microscopy data involves fitting the initial profile to a convolution of a Gaussian function with a Lorentzian function having a full-width at half-maximum of Γ, also known as a Voigt function. For one-dimensional diffusion, subsequent profiles can then be fit using Voigt profiles with the same Lorentzian component, since all diffusion dynamics and the change in the MSD will be captured in the increasing variance of the Gaussian component.[2,16] We refer to this procedure as the "Voigt fitting method".

Another approach is to avoid making any assumption about the initial profile shape, and simply use the experimentally measured profile at $t = 0$. In this method, the excited-state profile shapes are fit using a numerical convolution of the measured initial profile with the Gaussian expansion function.[29] The MSDs in this "numerical convolution method" are determined directly by the variance of the Gaussian expansion.

2.1.2. Direct numerical integration

An alternative method for extracting MSDs from time-resolved microscopy data is to abandon the profile-fitting framework of Eq. 5 entirely and obtain spatial variances by direct integration of the experimental data. In this "numerical integration method," the variances of the profiles at each time are directly calculated through the following sum:

$$\mathrm{Var}[n(\boldsymbol{x},t)] = \frac{\sum_i n(\boldsymbol{x}_i,t)\,(\boldsymbol{x}_i-\mu)^2}{\sum_i n(\boldsymbol{x}_i,t)} = \sum_i p(\boldsymbol{x}_i,t)\,(\boldsymbol{x}_i-\mu)^2 \tag{6}$$

Here, the sum is taken over all collected data points $\boldsymbol{x}_i$, where $p(\boldsymbol{x},t)$ is the excited-state probability distribution at time $t$. The MSDs are then determined by the change in variance from $t=0$. Conventionally, the reported MSDs correspond to transport along a single axis; if the variances were calculated from 2D or 3D data, they must first be divided by a factor of 2 or 3, respectively, to properly compare them to the 1D case. Unlike methods that fit the shape of profiles, the numerical integration method does not make any assumptions about the initial shape of the excited-state distribution or how it evolves over time.[31,32] For the ideal case of simple Brownian diffusion, the MSD extracted from Eq. 6 is theoretically the same as that obtained from Eq. 5.

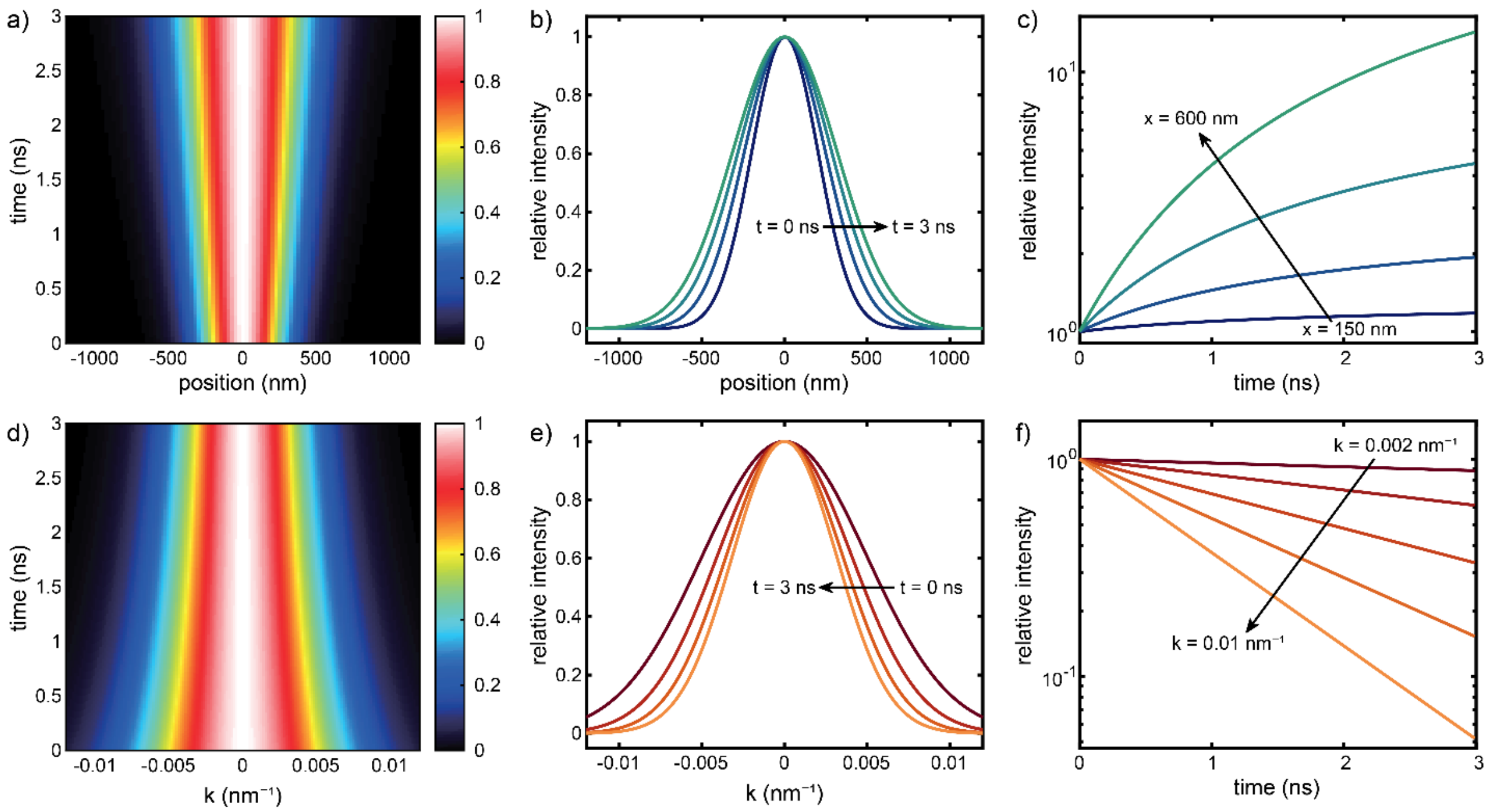

**Figure 1.** *Comparing real-space data representations to reciprocal-space.* a) Color map of the normalized profiles in real space (horizontal axis) and time (vertical axis) for a single diffusing species. b) Selected spatial profiles from panel (a) at t = 0 ns, 1 ns, 2 ns, and 3 ns. c) Selected temporal profiles from panel (a) at positions of x = 150 nm, 300 nm, 450 nm, and 750 nm, scaled to have the same intensity at t = 0 ns. d) Transformation of the data in panel (a) to reciprocal space (horizontal axis) and time (vertical axis). e) Reciprocal-space profiles from panel (d) at t = 0 ns, 1 ns, 2 ns, and 3 ns. f) Temporal profiles of selected normalized spatial modes from panel (d) at $k = 0.002$ $nm^{-1}$, 0.004 $nm^{-1}$, 0.006 $nm^{-1}$, 0.008 $nm^{-1}$, and 0.01 $nm^{-1}$. Each spatial mode is plotted relative to its intensity at t = 0 ns.

### *2.2. Reciprocal-Space Framework for Transient Microscopy Analysis*

As an alternative to existing MSD calculation techniques for transient microscopy, we propose analyzing the temporal evolution of spatial profiles in reciprocal space to extract transport parameters. This core mathematical concept has been successfully employed in related techniques such as dynamic light scattering, differential dynamic microscopy, and transient microscopy with structured excitation profiles, and it has numerous advantages that will be discussed throughout this work.[41–44]

The evolution of the excited species profile in reciprocal space can be determined by taking the Fourier transform $\mathcal{F}$ of the spatial profile $n(\boldsymbol{x},t)$, which is given by:

$$\mathcal{F}\{n(\boldsymbol{x},t)\} = \hat{n}(\boldsymbol{k},t) = \int n(\boldsymbol{x},t)\exp[-i\boldsymbol{k}\cdot\boldsymbol{x}]\,d\boldsymbol{x} \quad (7)$$

Converting the microscopy data to reciprocal space presents the excited-state dynamics from a new perspective. An expansion of the excited-state spatial profile over time in real space becomes a contraction over time in reciprocal space (Fig. 1). In reciprocal space, higher spatial frequencies correspond to intensity changes on shorter length scales. Since the action of diffusion is to smooth out spatial variations in intensity over time, the amplitudes of higher spatial frequencies will decay more rapidly than the amplitudes of lower ones (Fig. 1f). As we show later, analysis of the decay rates of different spatial modes can be especially useful for gaining insights into temporal dynamics.

In the ideal case of pure diffusive dynamics, the profiles will evolve according to the Fourier transform of Eq. 5:

$$\hat{n}(\boldsymbol{k},t) \propto \hat{n}(\boldsymbol{k},0)\exp[-\|\boldsymbol{k}\|^2 Dt] = \hat{n}(\boldsymbol{k},0)\exp\left[-\frac{\|\boldsymbol{k}\|^2 MSD}{2}\right] \quad (8)$$

The Fourier transform converts the convolution in real space from Eq. 5 to a simple multiplication in reciprocal space. Using Eq. 8, we can define a "Fourier fit method" in which the MSDs are determined by taking the Fourier transform of the spatial profile at each time, then fitting each profile to the product of the initial profile in reciprocal space and a Gaussian with variance $MSD^{-1}$.

### *2.3. Reciprocal-Space Methods for Cylindrical Geometry*

Although transport of excited species generally occurs in multiple directions within a material, transient microscopy data are often collected along a line or reduced to one dimension before the MSDs are calculated.[1,5,29,45] In some cases, this process can affect the accuracy of the MSD calculation methods outlined earlier. If the data are projected onto one dimension by integrating over other spatial dimensions, then the MSD calculation methods outlined in the previous sections are still valid. However, if the data are one-dimensional because they were collected along a line passing through the center of the excitation spot (which is frequently the case for transient photoluminescence microscopy) or obtained through azimuthal averaging of 2D datasets (which is frequently the case for wide-field transient scattering or transient absorption microscopy), then alternate MSD calculation techniques are needed. In these cases, the one-dimensional data effectively describe a radial profile of the excitation spot, and the evolution of the profile is most conveniently analyzed using cylindrical geometry. Even in cases where the initial profile is radially symmetric and evolves isotropically, convolution with a one-dimensional Gaussian spreading function generally fails to describe the evolution of the radial profile since it neglects the influence of transport perpendicular to the direction of the profile (see section 2.4.3). Mathematically, this means that:

$$\left(\iint n(x-\bar{x}, y-\bar{y}, 0) \exp\left[-\frac{\bar{x}^2+\bar{y}^2}{4Dt}\right] d\bar{x}d\bar{y}\right)\Bigg|_{y=0} \not\propto \left(\int n(x-\bar{x}, y, 0) \exp\left[-\frac{\bar{x}^2}{4Dt}\right] d\bar{x}\right)\Bigg|_{y=0} \tag{9}$$

To adapt the Fourier fit method for cases where the data describe radial profiles of the excited-state spatial distribution, one can employ Hankel transforms, which are defined as:

$$\mathcal{H}_n\{f(r)\} = \hat{f}(k) = \int_0^\infty f(r) J_n(kr)\, r\, dr \tag{10}$$

Here, $J_n$ is the $n$th-order Bessel function of the first kind. Notably, the $0^{th}$-order Hankel transform of a radially symmetric function is equivalent to the radial portion of a 2D Fourier transform up to a constant factor:

$$\mathcal{H}_0\{n(r,t)\} = \frac{1}{2\pi}\mathcal{F}_{2D}\{n(r,t)\} \tag{11}$$

As a result, Hankel transforms are a natural replacement for Fourier transforms when analyzing transient microscopy data in a cylindrical geometry. The equivalent of the convolution theorem for Hankel transforms is somewhat complex in its full form, but reduces to simpler expressions in many cases relevant to the transport of excited species. In particular, if $n(r,0)$ is a radially symmetric initial profile of some excited species, and if this profile expands isotropically according to some radially symmetric function $g(r,t)$, then the Hankel transform of the radial component of the profile can be written as:[46]

$$\mathcal{H}_0\{n(r,t)\} = \mathcal{H}_0\{n(r,0) * g(r,t)\} = \mathcal{H}_0\{n(r,0)\} \times \mathcal{H}_0\{g(r,t)\} \tag{12}$$

Here, $*$ denotes convolution over the full 2D plane, not just over the radial component. Eq. 12 also holds when $n(r,t)$ is the azimuthal average of an arbitrary 2D profile even if the full profile is not radially symmetric. Applying Eq. 12 to the Gaussian spreading function outlined in Eq. 5 yields:

$$\mathcal{H}_0\{n(r,t)\} \propto \mathcal{H}_0\{n(r,0)\} \times \exp[-k^2Dt] = \mathcal{H}_0\{n(r,0)\} \times \exp\left[-\frac{k^2MSD}{2}\right] \tag{13}$$

This is identical to the one-dimensional case of the equation used in Fourier fits (Eq. 8), except with the Fourier transform replaced by a Hankel transform. We can therefore define a "Hankel fit method", in which the MSDs are fit by numerically calculating the $0^{th}$-order Hankel transform of the azimuthally averaged spatial profile at each time, then fitting each profile to the product of the initial profile in reciprocal space and a Gaussian with variance $MSD^{-1}$.[47] For a more comprehensive overview of the relevant mathematics underlying Hankel transforms, and how it may be used to study anisotropic transport, see Baddour's work.[46]

| Technique | Initial Profile Fit Formula $n(x,0)$ | General Profile Fit Formula $n(x,t)$ | MSD Formula |
|---|---|---|---|
| Gaussian Fits | $\exp\left[-\frac{(x-\mu)^2}{2\sigma^2(0)}\right]$ | $\exp\left[-\frac{(x-\mu)^2}{2\sigma^2(t)}\right]$ with fixed mean $\mu$ | $\sigma^2(t)-\sigma^2(0)$ |
| Voigt Fits | $\frac{\Gamma_0/2}{x^2+\Gamma_0^2/4} * \exp\left[-\frac{(x-\mu)^2}{2\sigma^2(0)}\right]$ | $\frac{\Gamma_0/2}{x^2+\Gamma_0^2/4} * \exp\left[-\frac{(x-\mu)^2}{2\sigma^2(t)}\right]$ with fixed $\mu, \Gamma_0$ | $\sigma^2(t)-\sigma^2(0)$ |
| Numerical Convolution | Measured $n(x,0)$ | $n(x,0) * \exp\left[-\frac{x^2}{2\sigma^2(t)}\right]$ | $\sigma^2(t)$ |
| Numerical Integration (1D) | *N/A* | *N/A* | $\sum_i p(x_i,t)(x_i-\mu)^2 - \sum_i p(x_i,0)(x_i-\mu)^2$ |
| Numerical Integration (2D) | *N/A* | *N/A* | $\frac{1}{2}\left(\frac{\sum_i n(r_i,t)\, r_i^3}{\sum_i n(r_i,t)\, r_i} - \frac{\sum_i n(r_i,0)\, r_i^3}{\sum_i n(r_i,0)\, r_i}\right)$ |
| Fourier Fits | Measured $\hat{n}(k,0)$ | $\hat{n}(k,0) \times \exp\left[-\frac{k^2\sigma^2(t)}{2}\right]$ | $\sigma^2(t)$ |
| Hankel Fits | Measured $\mathcal{H}_0\{n(r,0)\}$ | $\mathcal{H}_0\{n(r,0)\} \times \exp\left[-\frac{k^2\sigma^2(t)}{2}\right]$ | $\sigma^2(t)$ |

**Table 1.** A summary of MSD calculation methods surveyed in this work. All profiles are additionally normalized as part of the fitting procedures.

### *2.4. Comparison of the Fourier Fit Method to Current State-of-the-Art*

To quantitatively compare the performance of different transient microscopy data analysis methods, we tested them side-by-side on simulated data under a variety of realistic experimental scenarios. First, we used finite difference methods to numerically solve Eq. 3 and create simulated transient microscopy datasets for the diffusion of an excited species in one dimension with defined diffusivity. To test the sensitivity of the calculation methods on the shape of the initial profile of the excited species, we used three different functions commonly encountered when analyzing microscopy data: a Gaussian function, a Voigt function, and an Airy disk (Fig. 2a). To determine the robustness of each MSD calculation method, we also added varying amounts of background noise and shot noise to our simulated data representative of typical data collected in our lab. We determined diffusivities from the MSDs calculated by each of the one-dimensional methods listed in Table 1 (Gaussian fits, Voigt fits, numerical convolution fits, numerical integration, and Fourier fits) to evaluate how accurately they could reproduce the true value of diffusivity used to generate the data. The full simulation workflow and parameters are described in section S1.

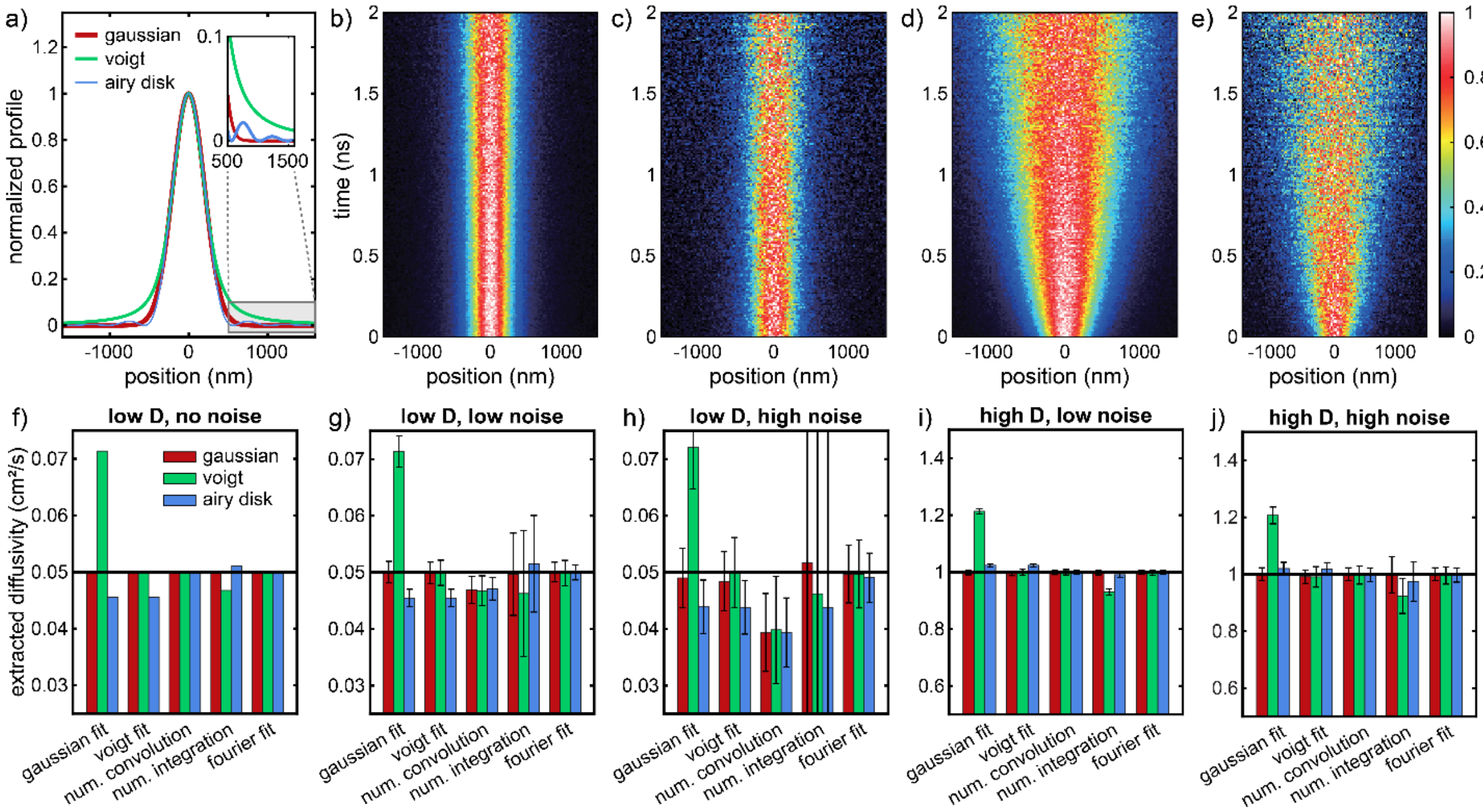

**Figure 2.** *Comparing the performance of the Fourier fitting method to conventional analysis methods.* a) Comparison of the Gaussian, Voigt, and Airy disk initial profile shapes used to create simulated datasets. Inset highlights the differences between the tails of these profiles. b-e) Simulated transient microscopy data sampling different diffusivity values and noise levels in Cartesian geometry: b) low-diffusivity, low-noise; c) low-diffusivity, high-noise; d) high-diffusivity, low-noise; e) high-diffusivity, high-noise. f-j) Extracted diffusivity values for each MSD calculation method (Gaussian fit, Voigt fit, numerical convolution, numerical integration, and Fourier fit) and initial profile shape (Gaussian, Voigt, and Airy disk from panel (a)) for Cartesian geometry. Bar heights represent the average extracted diffusivity value, and the error bars, when present, denote the standard deviation. The horizontal lines at 0.05 $cm^2/s$ (f-h) or 1 $cm^2/s$ (i-j) denote the true diffusivity value used in the simulation. The colors for each initial profile shape are consistent for all bar charts. Each panel represents a different combination of diffusivity value and noise level: f) low-diffusivity, no-noise; g) low-diffusivity, low-noise; h) low-diffusivity, high-noise; i) high-diffusivity, low-noise; j) high-diffusivity, high-noise.

### 2.4.1. Effect of initial profile shape

To identify systematic errors in the calculated diffusivity values derived from each of the MSD calculation methods, we first analyzed simulated transient microscopy data with no added noise (Fig. 2f). Overall, the Gaussian fitting method showed the most sensitivity to variations in the actual original profile shape. It was able to reproduce the correct diffusivity value when the initial profile was a Gaussian; but, when the assumption of a Gaussian initial profile shape was violated, the Gaussian fitting method calculated inaccurate diffusivity values that overestimated the true diffusivity by over 40% for the Voigt initial profile and underestimated the true diffusivity by about 9% for the Airy disk initial profile. Similarly, the Voigt fitting method returned accurate diffusivity values for the Gaussian and Voigt initial profiles, which can be exactly fit by Voigt functions, but underestimated the diffusivity values for the Airy disk initial profile. The numerical convolution method and Fourier fit method, which do not assume any initial profile shape, recovered the correct diffusivity values in all cases. Surprisingly, although the numerical integration method also makes no assumptions about the initial profile shape, it still calculated different diffusivity values for the various initial profile shapes, and was slightly inaccurate for the Voigt and Airy disk initial profiles (this effect is described further in section 2.4.5.).

2.4.2. Effect of signal-to-noise ratio

We explored the impact of the signal-to-noise ratio on the performance of each of the MSD calculation methods by adding two different amounts of noise to our simulated datasets. We added both shot noise and white noise to mimic sources of noise encountered in transient photoluminescence microscopy. The peak signal-to-noise ratio was ~30 and ~10 for the low- and high-noise cases, respectively (Fig. 2b, c). For both the low-noise and high-noise conditions, we calculated diffusivities for 100 iterations of random noise. The mean and standard deviation of the extracted diffusivity values for each calculation method and initial profile shape are shown in Figure 2g-h. The average accuracy of the Gaussian, Voigt, and Fourier fit methods was not significantly impacted by the addition of noise, and all three methods generally had comparable standard deviations under each combination of noise level and profile shape. Across all initial profile shapes, the standard deviation of the extracted diffusivity values for these three methods was around $0.001 - 0.003\, cm^2/s$, or 2-6% of the true value, in the low-noise case, and around $0.005 - 0.007\ cm^2/s$, or 10-14% of the true value, in the high-noise case. However, the addition of noise more severely affected the performance of the numerical convolution method, causing it to systematically underestimate the true diffusivity value of $0.05\ cm^2/s$ across all initial profile shapes. In the low-noise case, the mean diffusivity calculated by the numerical convolution method for each profile shape was around $0.047\ cm^2/s$. In the high-noise case, the mean diffusivity for each profile shape decreased further to around $0.04\ cm^2/s$. Overall, the precision of the numerical integration method was the most affected by the addition of random noise. The standard deviation of the diffusivity values calculated using numerical integration was around $0.007 - 0.011\ cm^2/s$ in the low-noise case, and increased to around $0.05 - 0.07\ cm^2/s$ in the high-noise case – larger than the diffusivity value itself. Out of the MSD calculation methods analyzed, only the Fourier fitting method was able to accurately calculate diffusivity values for all initial profiles while remaining relatively robust when random noise was added to the data.

The performance of all methods generally improved when the simulated diffusivity increased from $D = 0.05\ cm^2/s$ to $D = 1\ cm^2/s$ (Fig. 2d-e, 2i-j). This is expected intuitively, since the change in spatial profile width becomes more dramatic with increasing diffusivity. The diffusivity values extracted by the Gaussian fit and Voigt fit methods still had some dependence on the initial profile shape, but the relative magnitude of the error was less than in the $D = 0.05\ cm^2/s$ case. Additionally, the numerical convolution method no longer systematically underestimated diffusivity values under noisy conditions. Both this method and the Fourier fit method were able to accurately and precisely recover the correct diffusivity value for all initial profile shapes and noise levels. Finally, the larger diffusivity value improved the accuracy of the numerical integration method in the high-noise regime but had little effect on the relative accuracy in the low-noise regime. Across all methods, larger values of diffusivity improve the relative precision of the measurement (within reasonable limits).[30]

2.4.3. Effect of one-dimensional approximation to two-dimensional data

In general, one-dimensional models of transport do not accurately describe the evolution of radial profiles in cylindrically symmetrical systems, as the evolution of radial profiles is impacted by transport perpendicular to the radial axis (see section 2.3., Fig. 3a). To highlight the impact of the dimensionality of transport on the extracted diffusivity values, we repeated the quantitative assessment of the MSD calculation methods for simulated transport data in a cylindrical geometry with $D = 0.05\ cm^2/s$. We used the same parameters and simulation methods as in the one-dimensional case, while substituting the cylindrical Laplacian $\nabla^2 = \frac{1}{r}\frac{\partial}{\partial r}\left(r\frac{\partial}{\partial r}\right)$ for the Cartesian Laplacian $\nabla^2 = \frac{\partial^2}{\partial x^2}$. To enable more direct comparisons with the one-dimensional case and with typical formats of experimental data, we simulated the excited species concentrations along a line through the center of the profile, then averaged the two halves to obtain radial profiles when needed. We evaluated the performance of five MSD calculation methods on the simulated cylindrical data. Three of the methods – the numerical convolution, 1D numerical integration, and Fourier fit methods – are identical to the comparable methods used in the one-dimensional case described earlier, and are included here to demonstrate the limitations of using different 1D fitting methods on diffusion data in a cylindrical geometry. The fourth method we used was an adapted version of the 1D numerical integration method. For the case of the numerical integration method applied to radial profiles, one must include the cylindrical volume element $r\,dr$ in variance calculations:

$$MSD = \frac{1}{2}\left(\frac{\sum_i n(r_i,t)\, r_i^3}{\sum_i n(r_i,t)\, r_i} - \frac{\sum_i n(r_i,0)\, r_i^3}{\sum_i n(r_i,0)\, r_i}\right) \tag{14}$$

The factor of $1/2$ converts the MSD measurement to follow the convention of reporting transport parameters along a single direction. We refer to this modified equation as the 2D numerical integration method for calculating MSDs to distinguish it from the 1D version used earlier. Finally, we evaluated the performance of the Hankel fit method, as outlined in section 2.3.

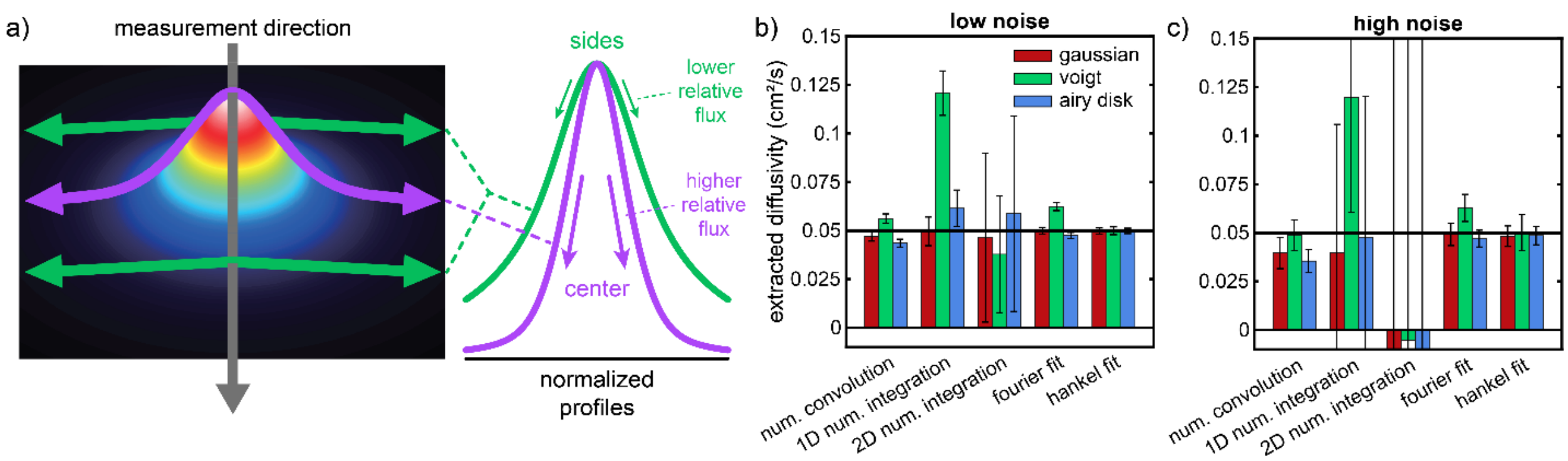

**Figure 3.** *Importance of correctly analyzing transient microscopy data in 1D.* a) Illustration of how changes in the profile curvature perpendicular to the measurement direction can impact the shape of the profile along the measurement direction. b-c) Extracted diffusivity values for select MSD calculation methods and varying initial profile shapes in cylindrical geometry under low-noise (b) and high-noise (c) conditions. Bar heights indicate the average extracted diffusivity value, and the error bars represent the standard deviation. The horizontal lines at 0.05 $cm^2/s$ denote the true diffusivity value used in the simulation.

The diffusivity values extracted by each of these calculation methods are given in Fig. 3b-c and Fig. S2. In general, none of the MSD calculation methods designed for 1D geometries (i.e., numerical convolution, 1D numerical integration, and Fourier fitting) were able to recover the correct diffusivity value for all profile shapes. In particular, the diffusivities extracted from the Voigt initial profile tended to be higher than the true diffusivity values, especially for the 1D numerical integration method and the Fourier fits. This is because the relatively sharper peak of the Voigt profile near the origin causes the profiles to flatten more rapidly than they would in a 1D geometry (Fig. 3a). On the other hand, the performance of the 1D analysis methods on Gaussian initial profiles was similar for both linear and cylindrical geometries, since cylindrically symmetrical Gaussian profiles are independent in the $x$- and $y$- axes and therefore evolve identically in both geometries. The 2D numerical integration method, despite being theoretically more accurate, failed dramatically in the case of high-noise data due to the amplification of uncertainty at the edges of the spatial distribution where signal-to-noise ratios are lowest. Finally, the Hankel fit method was accurate and relatively precise in both the low-noise and high-noise regimes across all initial profile shapes, making it the best-performing method overall for calculating MSDs in a cylindrical geometry.

2.4.4. Unique challenges of the numerical convolution fitting method

In the low-diffusivity regime with $D = 0.05\, cm^2/s$, the numerical convolution method systematically underestimated the diffusivity values when noise was added to the measurement (Fig. 2g, h). The main cause for this systematic underestimation is that the Gaussian function used in the numerical convolution must have a non-negative variance. This artificial lower bound can introduce an upward bias into the calculated MSD values, especially when the magnitude of the true MSD is comparable to the uncertainty in the MSD calculations caused by noise in the data (i.e. where the variance should cause the MSD to be a randomly distributed number around zero). Since MSDs are smaller at times closer to zero, the MSDs are primarily overestimated at early times, which decreases the slope of the MSD curves and the resulting diffusivity estimates. This effect is illustrated in Fig. 4a, which compares the average MSDs over time calculated by the Gaussian fit method and the numerical convolution method for a Gaussian profile in the high-noise regime. While the MSDs calculated by the Gaussian fit method generally follow the expected linear growth when $D = 0.05\, cm^2/s$, the MSDs calculated from the numerical convolution method are inflated at early times before converging to the values from the Gaussian fit method around $t = 1\,ns$, leading to underestimation of the true diffusivity.

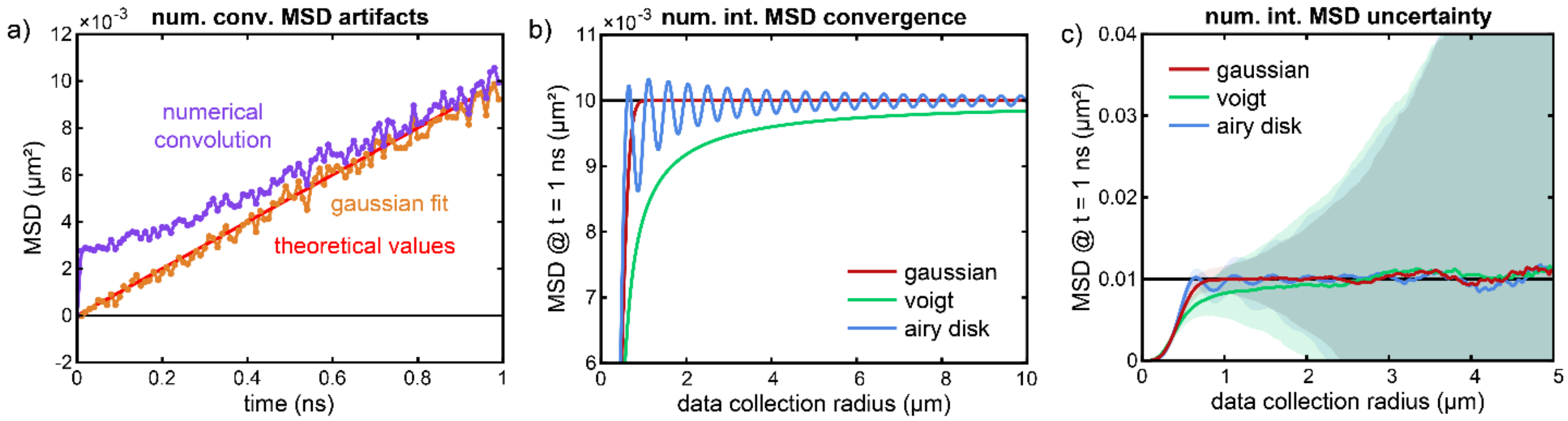

**Figure 4.** *Common artifacts in numerical convolution and integration methods.* a) Illustration of how numerical convolution can lead to underestimation of diffusivity on short timescales. b) MSDs calculated at t = 1 ns using the numerical integration method for different data collection radii and initial profile shapes in the low-diffusivity, no-noise regime. The black line indicates the theoretical value of 0.01 $\mu m^2$. c) Average MSDs calculated at t = 1 ns using the numerical integration method for different data collection radii and initial profile shapes in the low-diffusivity, low-noise regime. The shaded regions indicate the standard deviation of MSD calculations, and the black line marks the theoretical value of 0.01 $\mu m^2$.

To improve the performance of the numerical convolution method, the impacts of the overestimation of MSD values must be minimized. If the diffusivity is large enough, the magnitudes of the MSDs quickly grow larger than their uncertainties, limiting the extent to which MSDs are overestimated. For example, in our simulations with $D = 1\,cm^2/s$, the impact of MSD overestimation was negligible even in the high-noise regime, and the numerical convolution method was able to accurately recover the correct diffusivity value (Fig. 2i, j).

Alternatively, one can modify the analysis procedure to be able to recover negative MSD values by numerically convolving the profiles at time $t$ with a Gaussian of variance $\sigma_b^2$ before fitting them with Gaussian convolutions of the initial profile:

$$n(x,t) * \exp\left[-\frac{x^2}{2\sigma_b^2}\right] \propto n(x,0) * \exp\left[-\frac{x^2}{2\,MSD + 2\sigma_b^2}\right] = n(x,0) * \exp\left[-\frac{x^2}{2\sigma_c^2}\right] \qquad (15)$$

The added variance $\sigma_b^2$ can then be subtracted from the combined variance $\sigma_c^2$ to recover the MSDs. Since $\sigma_c^2$ may be less than $\sigma_b^2$, this method allows for negative MSD values.[48] We refer to this method as the "Gaussian blurring" method because one can understand the effect of convolving the initial measured profile with $\sigma_b^2$ as removing high-frequency noise from the initial profile shape. In theory, adding the blurring function [then later removing it] introduces no error for the case of normal diffusion (Gaussian spatial spreading). However, in practice, we find that choosing a magnitude for $\sigma_b^2$ systematically changes the recovered diffusivity in realistic datasets. Distortion of profile shapes introduced by the additional Gaussian blurring function can complicate data analysis if there are transport phenomena beyond the idealized, single-species diffusion simulated here.

2.4.5. Unique challenges of the numerical integration method

Although the numerical integration method makes the fewest assumptions about the behavior of the excited species in the material, it can struggle to accurately recover diffusivity values in certain cases. In our simulations, the numerical integration method did not recover accurate diffusivity values for Voigt and Airy disk profiles, even in the absence of noise (Fig. 2f). Unlike Gaussian distributions, these profiles have heavy power-law tails which can prevent the MSD estimate $\sum_i p(x_i, t)(x_i - \mu)^2 - \sum_i p(x_i, 0)(x_i - \mu)^2$ from converging over the finite range of $x$ values where data were collected. To illustrate this, we used the numerical integration method to calculate the MSDs of our low-diffusivity, no-noise simulations at $t = 1\ ns$ over different ranges of $x$ values $-r \leq x \leq r$, where $r$ is the data collection radius (Fig. 4b). While the MSD calculation for the Gaussian profile rapidly converged to the theoretical value of $0.01\ \mu m^2$, the MSD calculations for the Voigt and Airy disk profiles approached this value much more slowly, and they still had noticeable errors at the 2.5 $\mu m$ collection radius we used for our diffusivity calculations.

Although increasing the data collection radius can improve the accuracy of the numerical integration method, it also increases the sensitivity of the method to noise. The sum $\sum_i p(x_i, t)(x_i - \mu)^2$ evaluated in the numerical integration method amplifies noise in the tails of the profiles where $(x_i - \mu)$ is large, and increasing the data collection radius exacerbates this issue. To demonstrate this, we repeated the MSD calculations from Fig. 4b on the same data but under "low-noise" conditions where the peak signal-to-noise ratio (SNR) was 30. The resulting mean and standard deviations of the MSDs as a function of data collection radius are shown in Fig. 4c. Overall, the addition of noise caused the standard deviation of the MSD calculations to rise rapidly, becoming much greater than the magnitude of the MSD itself for data collection radii at just a few $\mu m$. The standard deviations had little dependence on the shape of the initial profile. In general, usage of the numerical integration method requires choosing a data collection range that balances capturing signal in the tails of the excited species profiles with limiting the impacts of noise on the MSD calculations, which may be challenging in many circumstances.

### *2.5. Effect of the instrument response function*

The temporal resolution of transient microscopy experiments is defined by the instrument response function (IRF). Depending on the setup, the IRF may be primarily determined by factors such as the laser pulse width or the response time of the photon detector.[4] Although all time-resolved spectroscopy experiments are affected by the IRF to some extent, its impacts on diffusivity estimates from transient microscopy measurements are not well understood. To explore these effects, we simulated transient photoluminescence microscopy data as above with $D = 0.5\ cm^2/s$, a Gaussian initial profile, and with lifetimes $\tau$ varying from 50 $ps$ to 1 $ns$. Following the simulations, we numerically convolved the data in time with a Gaussian IRF having a full-width at half-maximum of 100 $ps$, then re-defined $t = 0$ to be the instant with the maximum recorded signal. The resulting measured signal intensities for each lifetime are shown in Fig. 5a, and the corresponding MSDs (calculated using the Gaussian fit method) are shown in Fig. 5b.

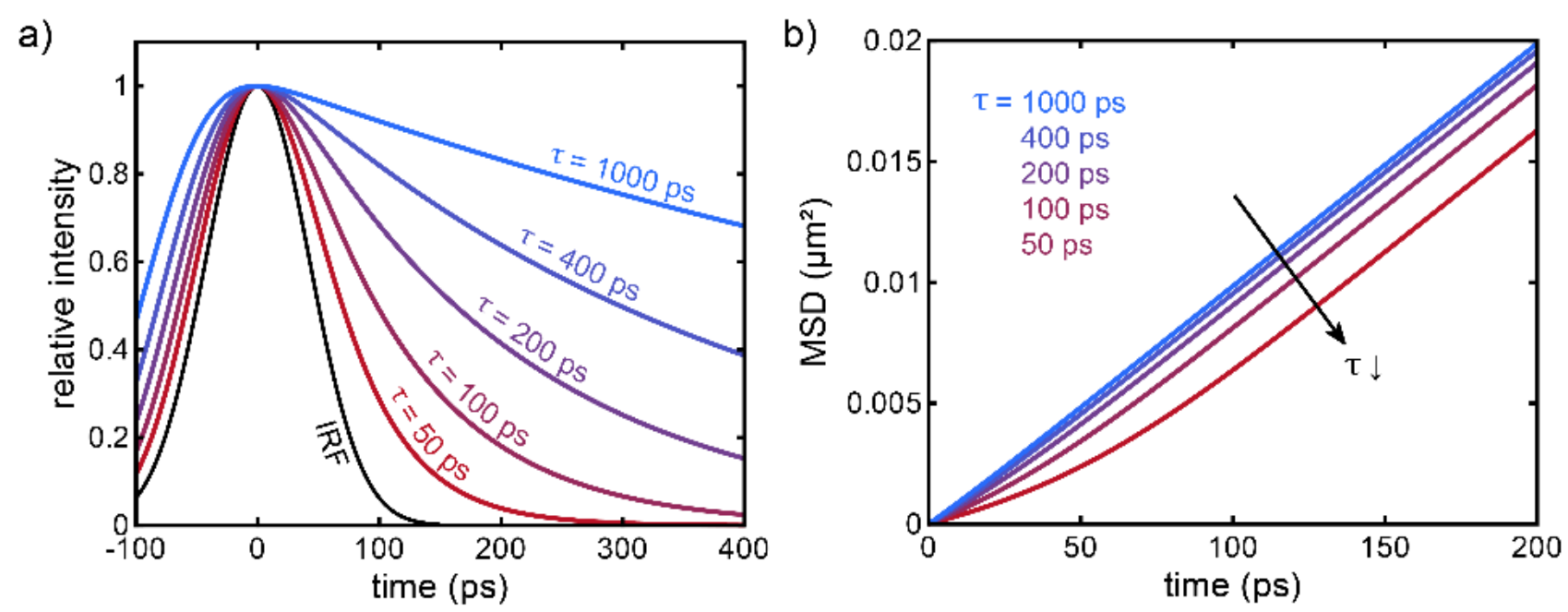

**Figure 5.** *Impacts of temporal instrument response function (IRF) on transient microscopy data.* a) Illustration of changes to the measured transient PL intensity decay as the lifetime ($\tau$) becomes comparable to the IRF (100 ps FWHM Gaussian IRF, in this example). b) Impact of the IRF on the MSD evolutions for species with different lifetimes. The MSD curves appear superlinear when the species lifetime is comparable to the IRF (100 ps, in this example).

When the excited-state lifetime was significantly longer than the 100 $ps$ time scale of the Gaussian IRF, the MSDs closely resembled the linear growth with a slope of $2D$ theoretically expected for diffusive behavior. However, as the lifetimes got shorter, the MSD curves began to grow more slowly and less linearly at early times. At the shortest lifetime of 50 $ps$, the growth of the MSD appeared noticeably superdiffusive and underestimated the theoretical MSD values by about 20% at 200 $ps$. We note that the use of alternate fitting methods described earlier does not substantially change the MSD curves in this case; all methods have similar artifacts introduced by the impact of the IRF.

These simulations demonstrate that when the IRF is comparable to the excited-state lifetime, it can introduce experimental artifacts that qualitatively and quantitatively distort the underlying dynamics at early times. Specifically, two systematic errors are introduced: 1) MSD curves appear *superlinear* when the excited-state lifetime is comparable to the IRF, and 2) the slope of the MSD curve at early times is shallower, leading to *underestimation* of the true diffusivity. Alternate analysis techniques, such as those which do not treat the measured profile at each time individually, are needed to account for the impacts of the IRF on the data.[18,49]

### 3. Classifying spatiotemporal dynamics using reciprocal-space analysis

Excited-state dynamics are generally much more complex than the idealized picture of a single diffusing species. Most of the MSD calculation methods described above assume a Gaussian profile expansion, which is usually not the case for anomalous transport phenomena. Additionally, MSD analysis alone is often insufficient for unraveling the changes in profile shape that result from complex excited-state dynamics.

To demonstrate the kind of complexity that can arise in physically realistic situations, we simulated excited-state dynamics in a simple system with two distinct species: a mobile species with a diffusivity of $D_1 = 1\ cm^2/s$ and an immobile species having a diffusivity of $D_2 = 0\ cm^2/s$. These two species could represent, for example, a population of band-edge [mobile] charge carriers co-existing with a population of trapped [immobile] charge carriers. In the simulation, an equal population of each species was initially excited at $t = 0$, and they had identical lifetimes of $\tau_1 = \tau_2 = 1\ ns$. The simulation did not allow for conversion/scattering between the mobile and immobile species, and the initial profile shape was Gaussian.

As shown in Fig. 6a, the resulting excited-state profile expanded in a non-Gaussian manner, becoming more heavy-tailed over time. Theoretically, the mean-squared displacement of the total population in this case is the average of the MSDs of the mobile and immobile species, and thus grows linearly over time with a slope representing the average diffusivity of the two species.

However, when we used the MSD calculation methods outlined earlier to analyze this dataset, only the numerical integration method was able to recover the correct MSDs, as the other methods were unable to correctly analyze the non-Gaussian profile expansion (Fig. 6b). Moreover, even the theoretically correct MSDs for this system are a poor diagnostic tool for uncovering the physical origins of the anomalous profile expansion, as they are exactly equivalent to the diffusion of a single species with $D = 0.5\ cm^2/s$. A more complete analysis of the changes in profile shape is needed to identify the presence of two different species in this case.

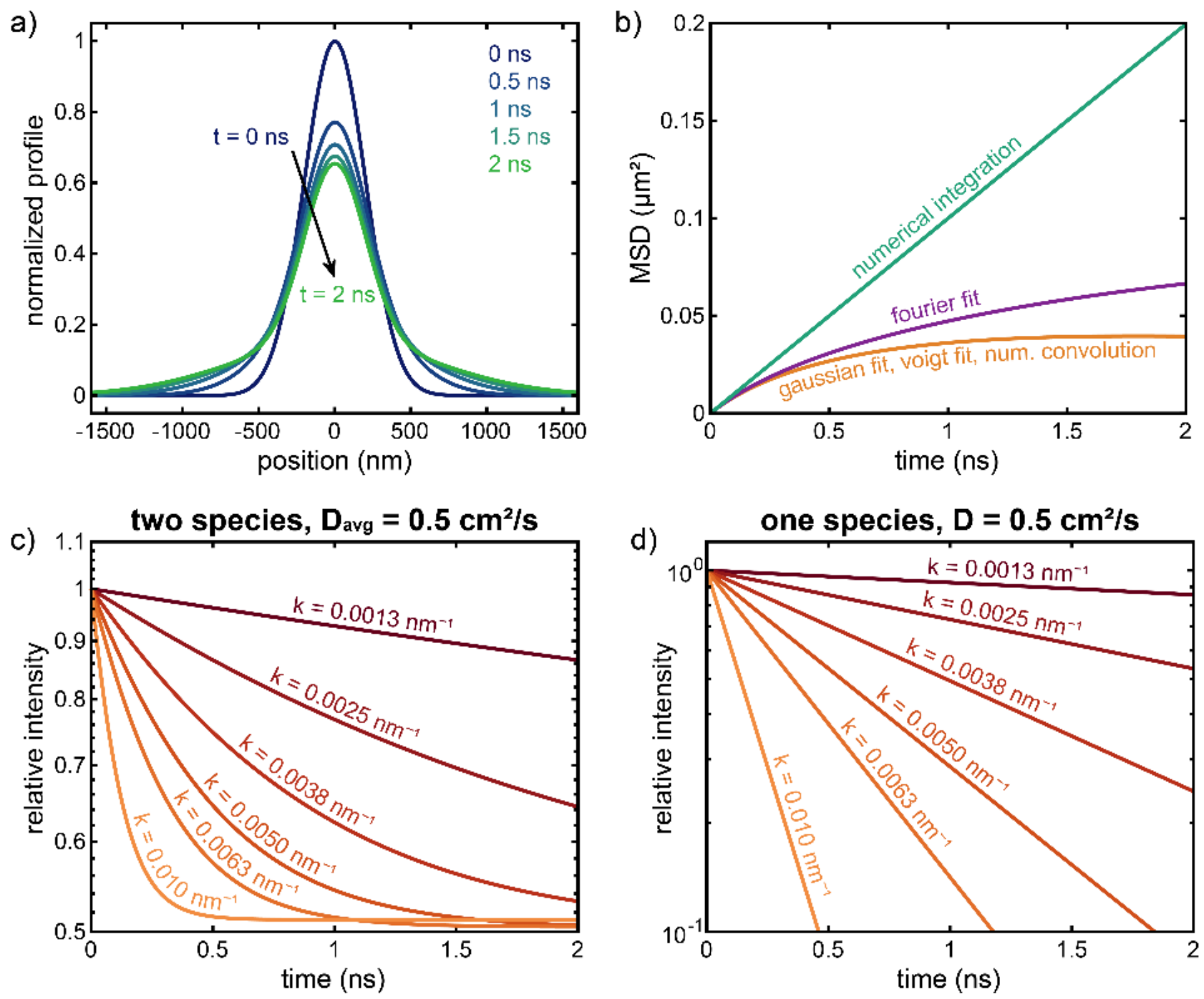

**Figure 6.** *Analysis of systems containing two diffusing species.* a) Area-normalized spatial profiles depicting the non-Gaussian evolution of a system containing two excited species with diffusivities of 0 $cm^2/s$ and 1 $cm^2/s$. b) Comparison of MSD curves for the two-species system depicted in panel (a), determined using different MSD calculation methods discussed previously. The Gaussian fit, Voigt fit, and numerical convolution methods converge to the same value. c) Normalized spatial mode decays for the system depicted in panels a-b with two diffusing species having diffusivities of 0 $cm^2/s$ and 1 $cm^2/s$. d) Normalized spatial mode decays for a single diffusing species with a diffusivity of D = 0.5 $cm^2/s$. Comparison of panels (c) and (d) reveals how spatial mode analysis can quickly reveal two-population behavior.

*3.1. A Fourier framework for analyzing complex dynamics*

To give a more complete and informative analysis of the temporal evolution of excited-state profiles, we propose analyzing the relative changes in the spatial mode intensities of normalized profiles over time, as in Fig. 1f. In these plots, the reciprocal space profiles at each time are normalized both spatially based on the amplitude for each $k$-value at $t = 0$, and temporally based on the amplitude at $k = 0$ for every point in time. This procedure results in normalized spatial mode intensities $f$ given by:

$$f(k,t) = \frac{\hat{n}(k,t)/\hat{n}(0,t)}{\hat{n}(k,0)/\hat{n}(0,0)} \tag{16}$$

Each point of $f(k,t)$ gives the normalized amplitude of a given spatial frequency at a specific time. By construction, $f(k,t)$ is constant at points where $k = 0$ or $t = 0$. Since the amplitude of the reciprocal space profiles at $k = 0$ corresponds to the total area under the real space curves, normalization by these values removes the overall decay dynamics and helps isolate the transport dynamics. The further normalization by the profile at $t = 0$ then limits or completely removes the impact of the original profile shape imposed by the laser excitation from the ensuing excited-state dynamics. The elegance of normalization in reciprocal space addresses a key weakness of the profile-fitting approaches in real space that was outlined earlier: although real-space curves are fit to the overall profile shapes, any information about transport is generally only contained in the *differences* between the profiles at different points in time. Even if real-space profiles are fit using an incorrect transport model, it is possible for the goodness of fit to be high simply because the model is based on the initial profile shape. This is especially true when there is little transport in the material, since in these cases the initial profile will have a greater influence on the shape of later profiles. Therefore, while real-space profile fitting can be an effective data analysis technique when the underlying transport mechanism is known, its ability to fully characterize transport is practically limited.

Normalized spatial mode analysis, and reciprocal space methods in general, provide a more efficient method of analyzing profile shapes. Although the full transformed profile may map onto a wide range of $k$-values, the information that can be transmitted at each spatial frequency is fundamentally limited by the optical transfer function of the microscope.[50] Therefore, one can restrict the analysis of $f(k,t)$ to only a few low-frequency $k$-values, removing the inherently noisy high-spatial-frequency modes from the analysis procedure while preserving information about the broad changes in profile shape. As a result, analysis of normalized spatial modes can give a concise and comprehensive overview of the profile evolution, allowing for a more informed analysis of the underlying transport phenomena. In the following sections, we will illustrate how several common physical phenomena that can be challenging to analyze using MSDs alone can be qualitatively and quantitatively analyzed using $f(k,t)$ and plots of normalized spatial mode decays.

### *3.2. Fourier analysis of transient microscopy data with multiple diffusing species*

There are commonly multiple species contributing to the signal in any transient microscopy experiment. These separate populations are often attributed to free and trapped charge carriers or excitons, but many systems have multiple species that move and decay at different rates.[33–39] We show here that normalized spatial mode decay plots (Fig. 1f, Fig. 6c-d) can be used to easily identify cases where one's transient microscopy data is influenced by multiple species with different transport parameters.

In the ideal case of a single diffusing species, the expression for the spatiotemporally normalized reciprocal space profiles $f(k,t)$ can be derived by normalizing both sides of Eq. 8 to yield the following:

$$f(k,t) = \exp[-k^2 Dt] \tag{17}$$

This expression indicates that for a single species, each normalized spatial mode undergoes exponential decay, with the decay rate given by the magnitude of $k$ and the diffusivity. Fig. 6d shows these simple diffusive dynamics for a single species with $D = 0.5\ cm^2/s$ using a log scale to depict the single-exponential decays as linear. However, the presence of multiple species with different diffusivities causes the spatial mode decay plots to become non-linear as they incorporate the different exponential decay contributions from each sub-population, similar to the effect of polydisperse particle distributions in dynamic light scattering experiments.[41] For example, in the model system introduced earlier that contains a mobile species with $D_1 = 1\ cm^2/s$ and an immobile species with $D_2 = 0\ cm^2/s$, the spatial mode decay plots become biexponential (Fig. 6c). Although the systems represented in Fig. 6c and Fig. 6d have identical MSD evolutions, the normalized spatial mode decay plots can clearly differentiate the dynamics occurring in each case. Relative to MSD analysis, normalized spatial mode decay analysis is far more capable of separating the diffusivities of different species based on their relative contributions to the decay dynamics at different $k$-values.

3.2.1. Negative effective diffusion

Different excited-state species often have varying lifetimes, exchange rates, and visibilities to the microscopy technique used. In many cases, the interactions between these terms can lead to the emergence of unusual transport behavior. One prominent example seen in many material systems is negative effective diffusion, where the MSD of the normalized data appears to *decrease* during some period of time. This behavior can emerge in systems characterized by a more mobile, short-lived species and a more stationary, long-lived species.[33,37,39,51,52] The population dynamics in these systems transition from being dominated by the mobile species at early times to being dominated by the less mobile, longer-lived species at later times, causing an apparent contraction in the profile (Fig. 7a). We emphasize that these apparent dynamics do not arise from a reversal in the direction of transport; rather, they arise from the disappearance of the short-lived, fast-diffusing species from the measurement ensemble. As before, standard MSD analysis can only give a partial picture of the full spatial profile evolution for this system (Fig. 7b).

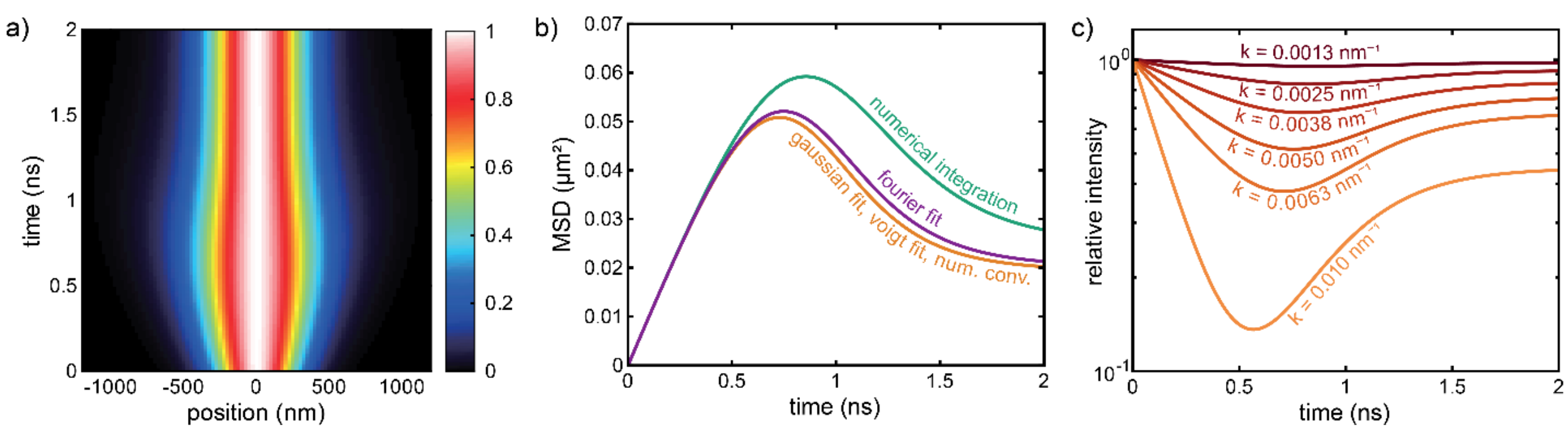

**Figure 7.** *Analysis of data exhibiting negative effective diffusion.* a) Normalized color map depicting the spatiotemporal evolution of a system with two species where the profile width contracts at later times. b) MSD for the system exhibiting negative effective diffusion depicted in panel (a), determined using each of the MSD calculation methods discussed previously. c) Normalized spatial modes for the system exhibiting negative effective diffusivity in panel (a).

Normalized spatial mode decay analysis can provide qualitative and quantitative insights into datasets exhibiting negative effective diffusion by leveraging all of the information contained in the profile shape (Fig. 7c). For example, we consider a general model with two different species, here labeled species 1 and species 2.[52] These two species have diffusivities of $D_1$ and $D_2$, and population decay rates of $k_{d1}$ and $k_{d2}$. We include interchange rates $k_{12}$ and $k_{21}$, respectively, for the conversion of species 1 to 2 and vice versa. The initial concentrations of species 1 and 2 are defined as $N_{1,init}$ and $N_{2,init}$, respectively. This system can be described by the following system of partial differential equations:

$$\frac{\partial n_1(x,t)}{\partial t} = D_1 \frac{\partial^2 n_1(x,t)}{\partial x^2} - (k_{d1} + k_{12})\, n_1(x,\ t) + k_{21}\, n_2(x,\ t) \tag{18a}$$

$$\frac{\partial n_2(x,t)}{\partial t} = D_2 \frac{\partial^2 n_2(x,t)}{\partial x^2} - (k_{d2} + k_{21})\, n_2(x,\ t) + k_{12}\, n_1(x,\ t) \tag{18b}$$

$$n_1(x,\ 0) = N_{1,init}\ \delta(x), n_2(x,\ 0) = N_{2,init}\delta(x) \tag{18c,d}$$

The magnitude of the signal in the microscopy data for a given concentration of species 1 or 2 is assumed to have a linear relationship with proportionality constants defined as $\Phi_1$ and $\Phi_2$, respectively (these parameters represent, for example, the photoluminescence quantum yield or absorption cross section of each species). Then, the total transient microscopy signal $q(x,t)$ will be:

$$q(x,\ t) = \Phi_1 n_1(x,t) + \Phi_2 n_2(x,t) \tag{19}$$

Following the approach of Adejumobi *et al.*,[52] the system specified by eqns. 18a-d can be readily solved in Fourier-Laplace space to yield the following equations for the transformed carrier distributions in terms of the spatial frequency *k* and Laplace variable *s*:

$$\hat{\hat{n}}_1(k,\ s) = \frac{C_2}{C_1 C_2 - k_{12}k_{21}} \left( N_{1,init} + \frac{N_{2,init}k_{21}}{C_2} \right) \tag{20a}$$

$$\hat{\hat{n}}_2(k,\ s) = \frac{C_1}{C_1 C_2 - k_{12}k_{21}} \left( N_{2,init} + \frac{N_{1,init}k_{12}}{C_1} \right) \tag{20b}$$

$$C_1 = s + D_1 k^2 + k_{12} + \ k_{d1} \tag{20c}$$

$$C_2 = s + D_2 k^2 + k_{21} + k_{d2} \tag{20d}$$

From here, an inverse Laplace transform can be performed to obtain an analytical expression for $\hat{q}(k,t)$. The function $\hat{q}(k,t)$ represents the impulse response function in Fourier space, which can be multiplied by the Fourier transform of the experimental spatial profile at *t* = 0 to obtain simulated experimental data. Due to its length, we give the equation for $\hat{q}(k,t)$ in section S2. Despite its complexity, the evaluation of this analytical solution in Fourier space is far more computationally efficient than numerically solving eqns. 18a-d. Fitting of the full data set in reciprocal space (Fig. 7c) is a more rigorous and information-rich alternative to simple MSD curve analysis (Fig. 7b) for identifying and quantifying the transport rates of different species in transient microscopy data.

### *3.3. Fourier analysis of anomalous transport phenomena*

Perhaps the most exciting scientific advance enabled by time-resolved microscopy is the ability to measure and characterize transport phenomena that deviate from Fick's Law. Based on the observed change in the MSD over time, anomalous transport behavior can be broadly classified as superdiffusive when the MSD grows superlinearly, and subdiffusive when the MSD grows sublinearly.[4] Many types of anomalous transport have been observed in a variety of material systems. For example, exciton hopping in disordered quantum dot solids appears subdiffusive at early times as the excitons thermalize within the energy landscape of the material.[2] Additionally, both charge carriers and quasiparticles such as excitons, polarons, ferrons, or magnons can undergo ballistic transport – movement with a consistent velocity – at certain length and time scales.[26,27,53–55] In particular, the ballistic movement of polaritons has attracted much attention in recent years due to their high speeds and long propagation lengths.[56–59] Furthermore, superdiffusive photon propagation has been observed in both atomic vapors and moderately doped semiconductors.[60–62] Finally, interactions between different excited species in transition metal dichalcogenides have been shown to lead to subdiffusive, superdiffusive, or even fluid-like transport under different conditions.[51,63–66] While a comprehensive overview of non-diffusive transport mechanisms is outside the scope of this work, in this section we examine a selection of non-diffusive phenomena seen in semiconducting materials to demonstrate how reciprocal-space analysis can give insights into the underlying transport mechanism.

3.3.1. Subdiffusive Transport

Subdiffusive transport in semiconductor materials is typically associated with energetic inhomogeneity. Aside from transitions to immobile trap states (explored in section 3.2.), subdiffusive behavior can also emerge in materials such as semiconducting polymers and quantum dot assemblies with significant inhomogeneous broadening.[1,2,4] In these materials, excited species transport through a series of hops among sites with different energies. As this happens, the average energy of the excited species decreases, which reduces the rate of subsequent hopping events since there are fewer energetically favorable transitions on average for lower-energy species. Eventually, the excited species reaches a thermal equilibrium over the energy landscape, and the transport behavior approaches steady state diffusion. Although the MSD curve in this case resembles subdiffusive behavior at early times, it is more accurate to describe this as non-equilibrium diffusive transport, with a time-dependent diffusivity that decreases towards a long-term, steady-state value as the excited species thermalize.[67] This mode of transport can be treated as a variant of the case of multiple diffusing species, but involving a continuum of states with different energies and diffusivities rather than a few discrete states as explored earlier.

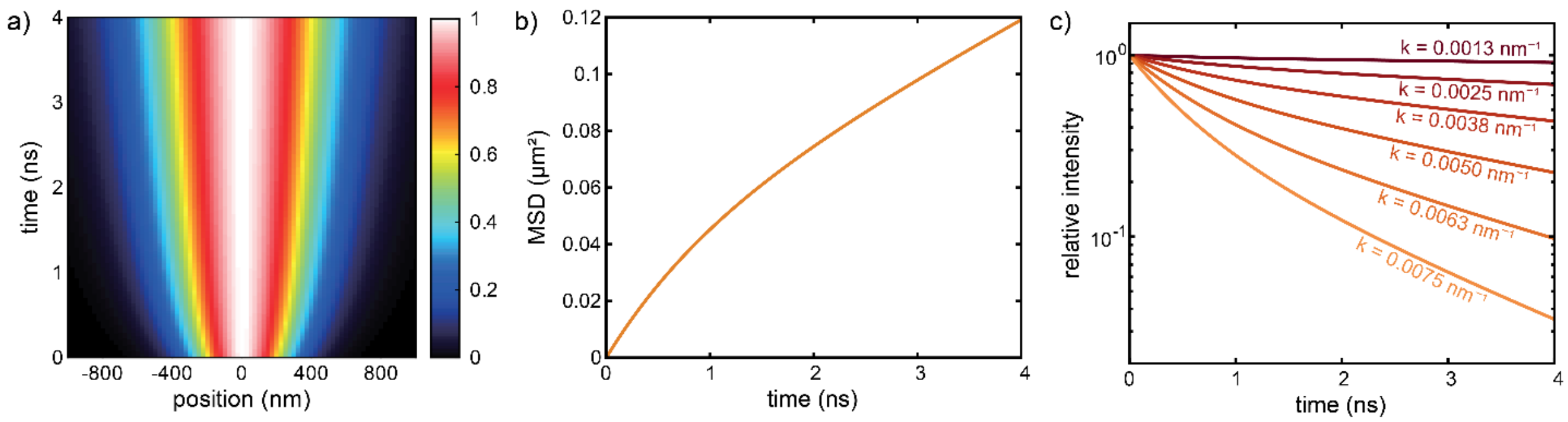

**Figure 8.** *Analysis of data appearing subdiffusive due to non-equilibrium population relaxation.* a) Normalized color map depicting the spatiotemporal evolution of a diffusing population with a temporally decaying diffusivity. b) MSD for the system exhibiting non-equilibrium diffusion depicted in panel (a), calculated using the Gaussian fit method. The MSD curve appears subdiffusive at early times due to the changes in diffusivity, but become linear (diffusive) at long times. c) Normalized spatial modes for the system exhibiting non-equilibrium diffusion in panel (a).

To show how non-equilibrium diffusive transport may appear in our reciprocal-space framework, we modified our finite difference simulations to solve the diffusion equation (Eq. 3) with a diffusivity value that decayed over time at a rate of $k_{relax} = 1\,ns^{-1}$ according to the following equation:

$$D(\,t) = (1 + 2 \cdot \exp[-k_{relax}t]) \times 0.1\frac{cm^2}{s} \tag{21}$$

The initial profile shapes and decay rate of $k_d = 1\,ns^{-1}$ were maintained from our previous simulations. The resulting spatiotemporal dynamics for the Gaussian initial profile are shown in Fig. 8a. The calculated MSDs in this case were similar regardless of fitting method. The MSD growth was sublinear at early times as the diffusivity decreased from its initial value of $D = 0.3\ cm^2/s$ to its equilibrium value of $D = 0.1\ cm^2/s$ (Fig. 8b). In reciprocal space, the decreasing diffusivity values were captured by the decreasing decay rates of each of the normalized spatial modes over the time scale of the changes in diffusivity (Fig. 8c). At later times, when the excited-state population had relaxed and the diffusivity had reached its equilibrium value, the MSD plot and the normalized spatial modes both transitioned to the linear behavior characteristic of standard diffusion (Fig. 8b-c).

3.3.2. Superdiffusive Transport – Ballistic Motion *vs* Photon Recycling

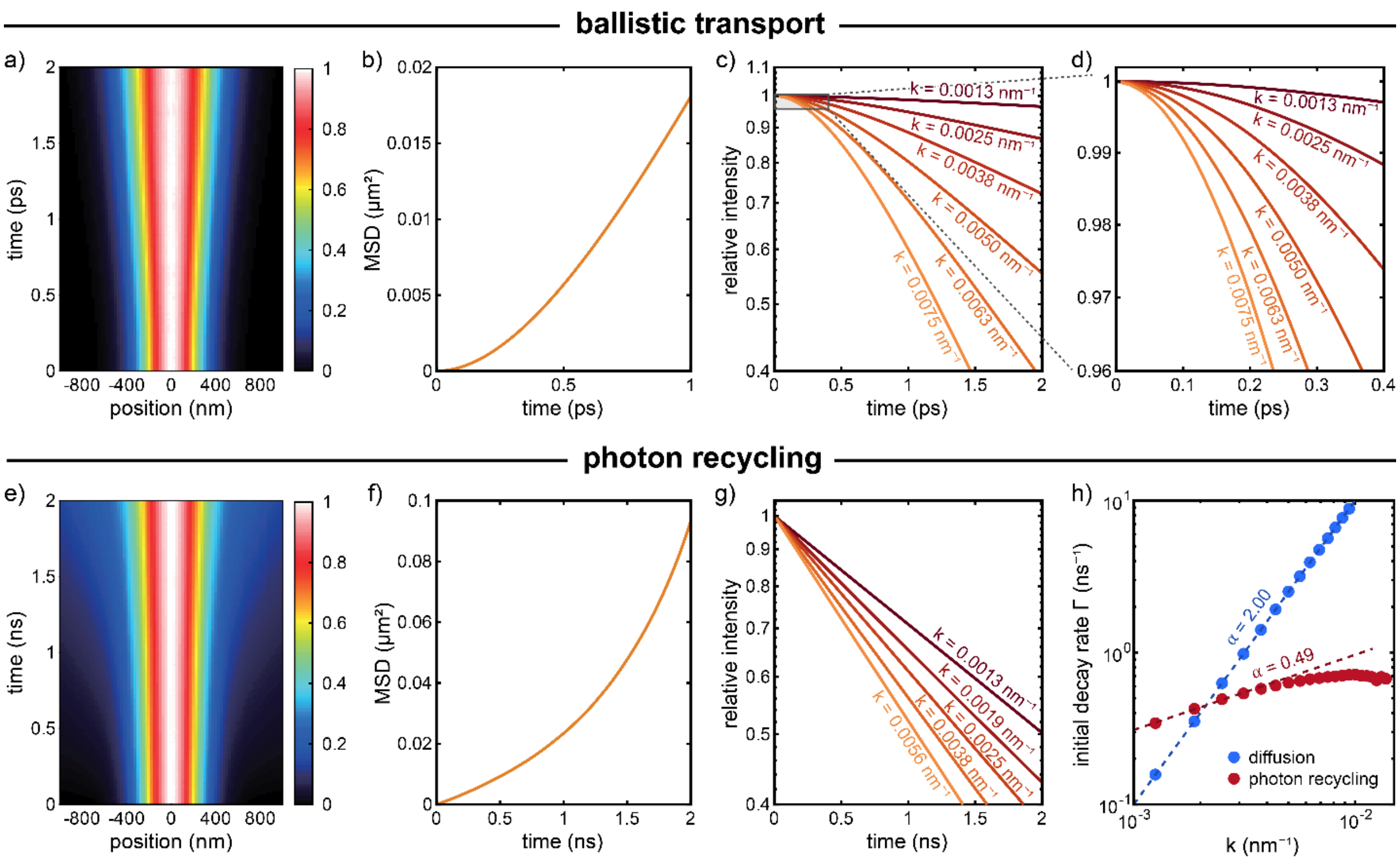

**Figure 9.** *Differentiating superdiffusive transport mechanisms by Fourier analysis: ballistic transport vs photon recycling.* a) Normalized color map depicting the spatiotemporal evolution of a population of charge carriers transitioning from ballistic transport at early times to diffusive transport at later times. b) MSD curve for the system undergoing ballistic transport depicted in panel (a), showing superlinear growth. c) Normalized spatial mode decays for the spatial profiles of charge carriers undergoing ballistic transport depicted in panel (a). d) Zoomed-in plot of the normalized spatial mode decays in panel (c), highlighting their quadratic behavior at early times. e) Normalized color map depicting the spatiotemporal evolution of concentration profiles for excitons undergoing transport by photon recycling (i.e. emission-reabsorption). f) MSD curve for the system undergoing photon recycling depicted in panel (e), showing superlinear growth. g) Normalized spatial mode decays for the excitons undergoing photon recycling depicted in panel (e). h) Initial decay rates of spatial modes as a function of *k* for systems undergoing transport via

diffusion (blue) and photon recycling (red). Dashed lines depict power-law fits to the decay rates at low values of $k$.

In bulk crystalline materials, charge carriers or excitons typically undergo “band-like” or “wave-like” transport, in which they move coherently until scattering off a phonon, crystalline defect, or other charge carrier. These scattering events alter the momentum of the particle, causing it to travel in a different direction until its next scattering event. This transport mechanism appears ballistic over time scales shorter than the characteristic scattering time but becomes diffusive over larger length and time scales as the impact of scattering events accumulates.

To illustrate how ballistic transport and the subsequent transition to diffusive behavior can be revealed by the normalized spatial decay modes, we created a kinetic Monte Carlo (KMC) simulation to model band-like transport of charge carriers. The positions of the charge carriers were initialized based on the Gaussian initial profile used earlier in the text. Each carrier traveled at a speed of $3 \times 10^5\ m/s$ in a random direction until it scattered into a new random direction. The scattering times were determined by sampling an exponential probability distribution with a mean of 500 fs, and the *x*-positions of the particles were recorded every 10 fs to create the final simulated dataset. The full details of the simulation are given in section S1.3.

The spatiotemporal evolution of the charge carriers in this discrete-scattering simulation is shown in Fig. 9a. As expected, the MSD curve for this simulation (calculated using the Gaussian fit method) exhibited superdiffusive behavior at early times (Fig. 9b) before transitioning to linear growth at later times. The normalized spatial modes, shown in Fig. 9c-d, also clearly illustrate the transition from ballistic to diffusive transport behavior through the changes in their curvature. The downwards curvature at early times captures the superdiffusive behavior at timescales up to the scattering time of the carriers. For this model of ballistic transport, the decay rate of each normalized spatial mode theoretically approaches 0 as $t \to 0$ (Fig. 9d, section S3). At later times, the normalized spatial modes become linear, indicating the transition to diffusive behavior. Normalized spatial mode analysis provides a method to both identify and quantitatively analyze ballistic motion (see section S3, Fig. S3).

Another possible transport mechanism that may appear superdiffusive is "photon recycling", or photon emission-reabsorption. In this mechanism, an excited species at one location in the material radiatively decays, and the resulting photon is absorbed at another location in the material to create a new excited species. The importance of photon recycling in energy transport in halide perovskites materials has received significant attention, and it likely plays an underappreciated role in many other semiconductor materials and devices.[40,68–71]

Photon recycling can be modeled as a continuous-time random walk with a step size probability distribution defined as $p(x)$. If the variance of the step size distribution and the mean time between steps are finite, then the transport will be diffusive at long times, and the displacement of the walkers after many steps will converge to a Gaussian.[60] For photon recycling, however, the step size distribution often has heavy, power-law tails that cause it to have infinite variance. This can occur when the emission spectrum overlaps with the tail of the absorption spectrum, resulting in large propagation lengths.[60–62,72] In these cases, the transport process is classified as a Lévy flight, and can be characterized by the parameter $\alpha$. The step size distribution for a Lévy flight has power-law tails according to the following relationship with $0 < \alpha < 2$:[73]

$$p(x) \propto |x|^{-1-\alpha} \tag{22}$$

Additionally, after many steps, the displacement of random walkers undergoing a Lévy flight will converge to a Lévy-stable density function in the long-time, long distance (small $k$) limit. This distribution can be defined in reciprocal space as:[73]

$$f(k,t) = \exp[-|k|^{\alpha}\lambda t] \tag{23}$$

Here, $\lambda$ parameterizes the overall rate of transport, analogous to the diffusivity in a diffusive system. Comparing Eq. 23 to Eq. 8, one can see that $f(k, t)$ is equivalent to the diffusive case with Gaussian spreading when $\alpha = 2$. However, despite this functional similarity, analysis techniques for diffusive transport can break down when applied to Lévy flights. Most notably, when $\alpha < 2$, the MSD diverges due to the heavy-tailed step-size distribution. Alternate analysis techniques are therefore required to quantitatively analyze Lévy flights.[73]

To explore how normalized spatial mode analysis could be used to help identify photon-recycling dominated transport in transient microscopy data, we created a KMC model to simulate photon recycling transport. We simulated excitons with radiative rates of $k_{rad} = 1\ ns^{-1}$ and with initial positions randomly assigned using the same Gaussian distribution as in our other simulations. Each exciton had a Gaussian photoluminescence spectrum and an absorption spectrum with a Gaussian red tail (Fig. S4). The Stokes shift of the excitons was set to be small so that the tail of the absorption spectrum would overlap with the emission spectrum. For each exciton, the step durations were randomly determined from an exponential distribution defined by $k_{rad}$. At each step, the propagation length was determined by randomly sampling a wavelength from the photoluminescence spectrum, determining the absorptivity at that wavelength, and using the Beer-Lambert law to randomly select a propagation distance for that absorptivity. The direction of emission was randomized, and the *x*-position of each exciton was recorded every 10 ps to construct the final dataset. The full simulation details are given in section S1.3.

Simulated time-resolved microscopy data in the case of photon-recycling-dominated transport is shown in Fig. 9e. Although the MSDs are formally undefined, the MSDs calculated by the Gaussian fit method appeared superlinear for this system, similar to the ballistic transport case (Fig. 9f). However, unlike the ballistic transport case – where the initial slope of spatial modes approaches zero at early times – the normalized spatial modes for a Lévy flight appear linear, indicating exponential decay similar to diffusive transport (Fig. 9g). Determining whether the excitons in our simulation are undergoing diffusion or a Lévy flight requires calculating the value of $\alpha$ for this transport. To do this, we measured the decay rate Γ of the spatial modes at each $k$. The decay rates as a function of $k$ should follow a power law with an exponent of $\alpha$ according to:

$$\Gamma(k) = |k|^{\alpha}\lambda \tag{24}$$

We fit Eq. 24 to the decay rates Γ for our photon recycling simulation, discarding the smallest non-zero value of $k$ to limit windowing artifacts from the finite size of our dataset (see Fig. S5). For diffusion, the decay rate should have a quadratic relationship with $k$ ($\alpha = 2$). However, Fourier analysis of our photon recycling simulation returned a much smaller value of $\alpha = 0.49$ in the small $k$ limit, indicating that the exciton transport in this case is better classified as a Lévy flight (Fig. 9h, also see section S4).

We emphasize the value of spatial mode analysis in differentiating transport processes – especially an understanding of spatial mode decay rate Γ *vs* spatial frequency $k$, as shown in Figure 9h. For example, normalized spatial mode decays may be non-linear due to the presence of multiple diffusing species or due to temporal variations in the diffusivity, as shown in Figs. 6c, 7c, & 8c. However, the decay rates at early times in these cases will still follow a quadratic relationship with $k$ due to the underlying diffusive nature of the transport (see section S5).

Overall, these simulations of ballistic transport and photon recycling illustrate the high flexibility of normalized spatial mode analysis and the ways it can be used to gain insights about transport phenomena beyond simple diffusion. Normalized spatial mode analysis can also give insights beyond MSDs for identifying certain transport mechanisms. For example, ballistic transport and photon recycling as outlined above may be challenging to distinguish using MSD analysis alone since they are both superdiffusive mechanisms associated with superlinear MSD growth (Fig. 9b,f). However, they are governed by fundamentally different transport dynamics which can be identified by a more careful look at the full reciprocal-space profile evolution. In general, different transport mechanisms often have key features in their reciprocal space dynamics which can be used to gain qualitative and quantitative insights into the underlying material physics.

### *3.4. Spatially varying recombination rates & non-linear interactions*

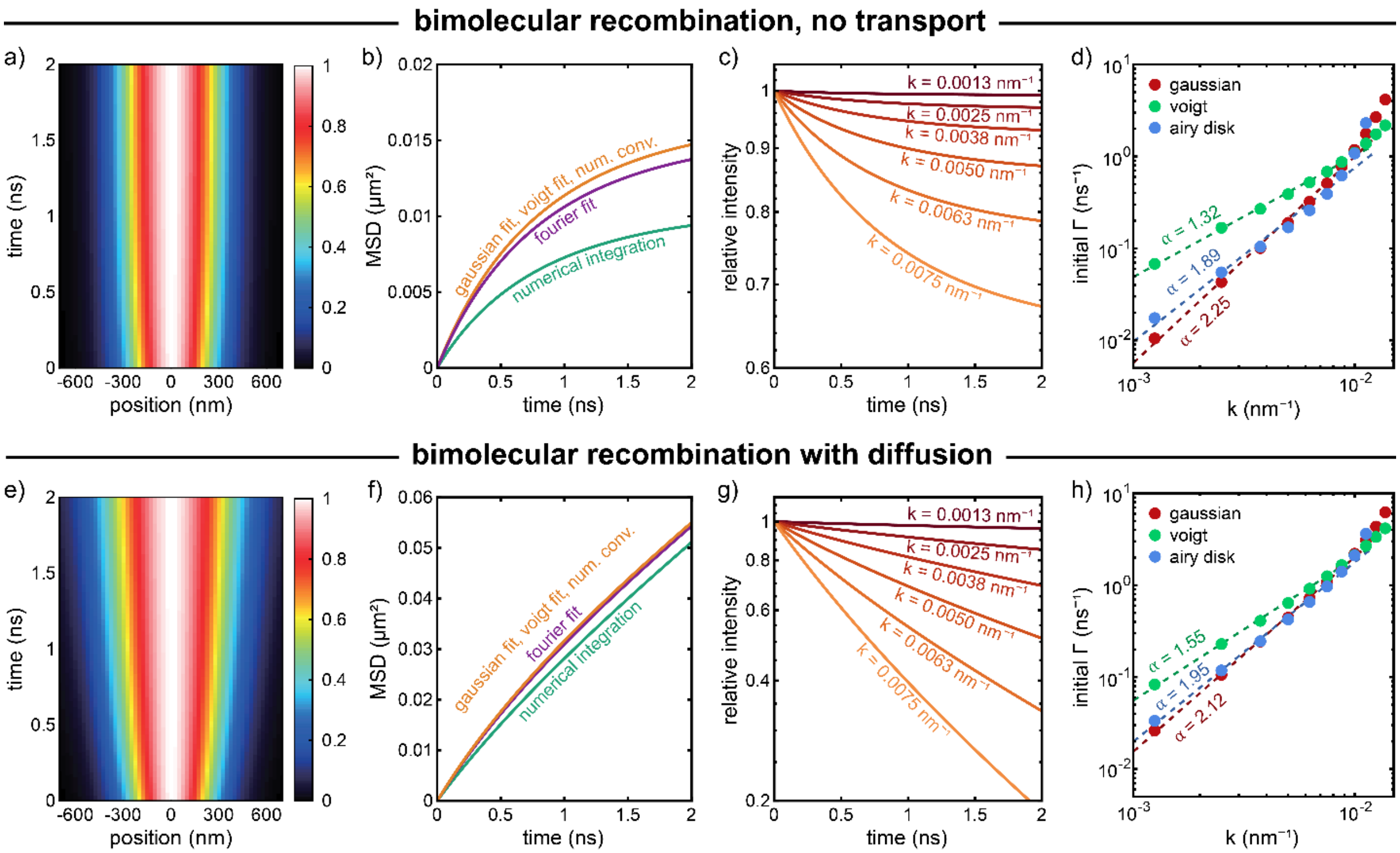

**Figure 10.** *Analysis of systems exhibiting bimolecular recombination*. a) Normalized color map depicting the spatiotemporal evolution of concentration profiles for a stationary species with both unimolecular and bimolecular decay terms. b) MSD for the stationary species with both unimolecular and bimolecular decay terms depicted in panel (a), exhibiting sublinear growth. c) Normalized spatial mode decays for the stationary species with non-linear decay terms depicted in panel (a). d) Initial decay rates of spatial modes within the first 50 ps as a function of $k$ for simulations of stationary species with non-linear decay terms and different initial profile shapes. Dashed lines depict power-law fits to the decay rates for each initial profile shape at low values of $k$. e) Normalized color map depicting the spatiotemporal evolution of concentration profiles for a species with unimolecular decay, bimolecular decay, and diffusion terms with $D = 0.1\ cm^2/s$ . f) MSD for the species undergoing diffusion and non-linear decay depicted in panel (e), exhibiting

sublinear growth to a lesser extent than the stationary species in panels (a-d). g) Normalized spatial mode decays for the diffusing system with non-linear decay terms depicted in panel (e). h) Initial decay rates of spatial modes within the first 50 ps as a function of $k$ for diffusing species with non-linear decay terms and different initial profile shapes. Dashed lines depict power-law fits to the decay rates for each initial profile shape at low values of $k$.

Another common complication in the analysis of transient microscopy data is the presence of non-linear decay mechanisms, such as bimolecular charge carrier recombination, Auger recombination, or triplet-triplet annihilation.[49,74] Since these processes selectively increase the decay rate in regions with higher concentrations, they will cause the normalized spatial profiles of the excited-state species to broaden over time, even in the absence of any transport. The broadening effect that non-linear recombination has on profiles in real space and the resulting MSD analysis has been covered in prior works.[17,18,20,40] In this section, we demonstrate the effect these dynamics have on reciprocal-space analysis of transient microscopy data.

When the spatial evolution of an excited species is independent of its initial spatial distribution, normalizing to the initial profile is a powerful tool that isolates the key physics occurring in the system. Formally, this is equivalent to systems where, regardless of the shape of the initial profile $n(\boldsymbol{x}, 0)$, the measured signal $n(\boldsymbol{x}, t)$ at any time can be expressed as a convolution between the initial profile and a given function $u(\boldsymbol{x}, t)$ capturing the dynamics of the system:[29]

$$n(\boldsymbol{x}, t) \propto n(\boldsymbol{x}, 0) * u(\boldsymbol{x}, t) \tag{25}$$

This is a flexible formulation that can be adapted to many types of transport including normal-, sub-, and super-diffusive behavior, and analysis of co-diffusing species.

However, Eq. 25 is invalid for systems with non-linear interactions or spatially varying parameters. In these cases, the initial shape of the profile plays a key role in the evolution of the system, and must be accounted for in any analysis of the spatiotemporal dynamics. As we show later in this section, depending on the shape of the initial profile, the normalized spatial modes of these systems may share certain features with various forms of anomalous transport. Any assignment of anomalous transport behavior based on the normalized spatial mode analysis techniques presented here should therefore be done cautiously. The best way to diagnose such non-linear dynamics is by 1) experimentally determining if the MSD curve depends on excitation laser power or repetition rate, and 2) assessing which term – diffusion or recombination – dominates the spatial broadening on the timescale of the experiment (see Nagaya Wong *et al.*[18] for detailed analysis).

To explore how the normalized spatial mode analysis is affected by non-linear recombination, we performed one-dimensional simulations of the evolution of profiles of excited-state species undergoing diffusion with diffusivity $D$, unimolecular decay with a rate constant of $k_d$, and bimolecular decay with a rate constant of $k_{bi}$. This system is governed by the following differential equation:

$$\frac{\partial n(\boldsymbol{x},t)}{\partial t} = D\nabla^2 n(\boldsymbol{x},t) - k_d\, n(\boldsymbol{x},t) - k_{bi}\, n(\boldsymbol{x},t)^2 \qquad (26)$$

Using our finite difference simulation framework from before, we simulated one-dimensional systems following Eq. 26 and using the Gaussian, Voigt, and Airy disk initial profiles given in Fig. 2a. To establish a baseline for how nonlinear recombination affects profile shapes in reciprocal space, we initially simulated the dynamics of an immobile species (*i.e.*, with $D = 0\ cm^2/s$) undergoing unimolecular and bimolecular decay with rate constants of $k_d = 1\ ns^{-1}$ and $k_{bi} = 1\ nm^3/ns$, respectively. The peak concentration of each initial profile was set to $1\ nm^{-3}$.

Fig. 10a-d shows the results of these spatiotemporal simulations including bimolecular recombination with no diffusion. The spatial profiles for every initial distribution appeared to broaden, as represented in Fig. 10a for the Gaussian initial profile. This led to MSD growth at early times. However, as the population decayed, the relative impact of the bimolecular decay term

decreased, slowing the profile expansion and causing the MSD curve to become sub-linear (Fig. 10b).[40] Additionally, the normalized spatial modes of each initial profile decayed non-linearly (Fig. 10c). The decay rate of each spatial mode decreased over time, similar to the behavior of two diffusing species or nonequilibrium diffusion seen in Figs. 6c and 8c, respectively. Additionally, the initial decay rates for the normalized spatial modes over the first 50 ps (Fig. 10d) deviated from the quadratic behavior expected for standard diffusion, similar to the decay rate plots for Lévy flights (Fig. 9h). The decay rates additionally became noticeably non-linear at higher values of $k$. However, unlike the transport mechanisms discussed earlier (Figs. S6-S9), the behavior of the normalized spatial mode decay plots was not characteristic of the type of dynamics occurring in the system, but *varied* according to the shape of the initial profile (Figs. 10d, S10). As shown in Fig. 10d, the value of the exponent $\alpha$ in Eq. 24 that best fit the spatial mode decay rates at lower values of $k$ was different for each initial profile shape, ranging from $\alpha = 1.32$ for the Voigt initial profile to $\alpha = 2.25$ for the Gaussian initial profile.

Fig. 10e-h shows the effect of reintroducing diffusive dynamics into our simulations of a species undergoing bimolecular decay. We chose to simulate a diffusivity of $D = 0.1\ cm^2/s$ so that the increases in profile width from both bimolecular recombination and diffusion would be similar at early times. The resulting spatiotemporal dynamics, shown in Fig. 10e for the Gaussian initial profile, combined key features from both of these mechanisms. The MSD growth and normalized spatial mode decays remained non-linear and continued to appear subdiffusive (Figs. 10f-g), although the deviations from linearity were less dramatic than in the case with $D = 0\ cm^2/s$. Additionally, the initial decay rates of the normalized spatial modes as a function of $k$, fit using Eq. 24, still varied based on the initial profile shape and deviated from the quadratic relationship with $\alpha = 2$ that is characteristic of pure diffusion (Figs. 10h, S11). However, for each initial profile shape, the value of $\alpha$ was closer to 2 than in the corresponding simulation with $D = 0\ cm^2/s$, reflecting the influence of the diffusive term on the changes in the spatial profiles.

## 4. Conclusions

In this work, we evaluated a wide variety of techniques for transient microscopy data analysis. We first simulated transient microscopy datasets with a range of transport parameters and commonly encountered experimental artifacts to evaluate the quantitative performance of different MSD calculation methods. We highlight several problems that conventional analysis techniques may face under different conditions:

- Calculation techniques which fit the profile to a specific functional form (*e.g.*, the Gaussian fit and Voigt fit methods) are generally inaccurate when the measured profile deviates from the assumed curve shape.
- Calculating MSDs using numerical convolutions of the initial profile can systematically underestimate diffusivity values, especially when the data is noisy and the underlying diffusivity is small. This issue can theoretically be mitigated using techniques like Gaussian blurring.
- The "numerical integration method", in which MSDs are calculated by using numerical integration to estimate the variances of profiles, is more susceptible to noise than other calculation methods.
- Even in the absence of noise, the variance estimates from the numerical integration method do not necessarily converge within the width of typical spatiotemporal microscopy datasets, leading to inaccurate diffusivity estimates.

- If microscopy data is collected by measuring along a line passing through the center of the excitation spot, or are projected onto a radial profile, then the one-dimensional Gaussian spreading framework underlying many MSD calculation methods is generally inaccurate. In these cases, one must account for the full two-dimensional shape of the imaged excited-state profile to determine how it will evolve.
- When the excited-state lifetime is comparable to the temporal IRF, the effects of the IRF will distort the MSD curve at early times. Specifically, the MSDs will appear superlinear and diffusivity will be underestimated on the time scales of the IRF.
- Regardless of how it is calculated, mean-squared displacement (MSD) is a summary statistic with a limited ability to describe the nuances of how the spatial profiles of excited species evolve. As a result, MSD analysis can struggle to identify and characterize complex transport behaviors.

To resolve these issues, we proposed a series of analysis techniques based on analyzing excited-state spatial profiles in reciprocal space. First, we introduced the Fourier and Hankel fit methods for MSD calculation. These methods resolve many of the issues with conventional data analysis techniques listed above: they make no assumptions about the initial profile shape, they are less sensitive to noise than the numerical convolution and numerical integration methods, and together they account for both Cartesian and cylindrical geometries in data set collection. Through our simulations, we demonstrated that the Fourier and Hankel fit methods have the best overall performance in extracting the correct diffusivity value in Cartesian and cylindrical geometries, respectively.

We also extended our reciprocal-space-based techniques beyond MSD calculation, outlining a method that gives a more comprehensive view of the spatial evolution of the excited species distribution through analyzing changes in the normalized spatial modes of the profile shapes. We demonstrated how this method can be used to help identify and quantitatively analyze various transport phenomena beyond single-species diffusion, including diffusion of multiple species, subdiffusive behavior, ballistic transport, and photon recycling. However, further refinements are required to accurately analyze data with non-linear dynamics or complex spatial variations in decay or diffusion parameters, and care must be taken to differentiate these phenomena from the various transport mechanisms which reciprocal-space analysis is well suited to identify and analyze. These reciprocal-space analysis methods have the potential to both improve the quantitative accuracy of parameter estimates and to provide a more complete picture of the transport dynamics, giving further insights into the wide range of excited-state behaviors in materials.

ASSOCIATED CONTENT

**Supporting Information**.

Our transient microscopy reciprocal-space analysis toolbox is available at https://github.com/tisdalelab/transient_microscopy_reciprocal_space_analysis_toolkit. This repository contains both MATLAB and Python implementations of the analysis techniques discussed in the paper.

Supplementary Information (PDF)

- Section S1. Simulation details
    - S1.1: Initial profile shapes
    - S1.2: Finite difference simulations
    - S1.3: KMC simulations
- Section S2: Analytical solution to two-species diffusion with interchange
- Section S3: Quantitative analysis of ballistic transport
- Section S4: Decay rate scaling for Lévy flights in reciprocal space
- Section S5: Early-time behavior of 2 diffusing species in reciprocal space
- Section S6: Supplementary Figures S1-S11

**Acknowledgements**

This work was supported by the U.S. Department of Energy, Office of Science, Basic Energy Sciences, under award no. DE-SC0019345. T.J.S. was partially supported by an MIT Chemical Engineering Mathworks Fellowship.

**Conflict of Interest Statement**

The authors have no conflicts to disclose.

**Author Contributions**

Thomas J. Sheehan: conceptualization (lead), data curation (lead), formal analysis (lead), investigation (lead), methodology (lead), software (lead), validation (lead), visualization (lead), writing/original draft preparation (lead), writing/review and editing (lead). William A. Tisdale: conceptualization (supporting), funding acquisition (lead), project administration (lead), resources (lead), supervision (lead), writing/original draft preparation (supporting), writing/review and editing (supporting).

**Data Availability Statement**

Upon acceptance, the data that support the findings of this study will be made available at https://github.com/tisdalelab/transient_microscopy_reciprocal_space_data

Supplementary Information for:

# Analysis of Transient Microscopy Data in Reciprocal Space

*Thomas J. Sheehan*[1] *and William A. Tisdale*[1]*

Author Addresses:

[1]Department of Chemical Engineering, Massachusetts Institute of Technology, Cambridge, Massachusetts 02139, United States

<u>Contents</u>

## S1. Simulation details

### S1.1. Initial profile shapes

The following functions were used as the initial profile shapes for the simulations in this text. These profiles were scaled by varying constant factors as appropriate for different parts of the text. Here, the $*$ symbol denotes convolution, and $J_1$ is the first-order Bessel function of the first kind.

| | | |
|---|---|---|
| Gaussian | $n(x,0) = \exp\left[-\frac{x^2}{2 \cdot 200^2}\right]$ | (S1a) |
| Voigt | $n(x,0) = \exp\left[-\frac{x^2}{2 \cdot 135^2}\right] * \frac{1}{125^2 + x^2}$ | (S1b) |
| Airy Disk | $n(x,0) = \left(\frac{J_1(6.844 \times 10^{-3}\, x)}{x}\right)^2$ | (S1c) |

*Table S1. Functional forms of initial profile shapes.*

### S1.2. Finite difference simulations

For each simulation, the evolution of the spatial profiles was calculated by numerically solving the relevant partial differential equation or equations using finite difference approximations. Second-order centered difference approximations were used to estimate the spatial derivatives, and temporal changes in the profile were estimated using the explicit Euler method. The signal at each time $q(x,\, t)$ was modeled using:

$$q(x,\, t) = \sum_i \Phi_i n_i(x,t) \tag{S2}$$

Here, the sum is taken over all species contributing to the signal, each with a proportionality constant $\Phi_i$. Once the simulations were complete, the results were binned spatially and temporally and truncated to match more realistic data acquisition conditions. The physical and simulation parameters used throughout this text are summarized in tables S2 and S3, respectively.

| **Parameter** | Single-species diffusion | Two species diffusion, no interchange | Two species diffusion, negative apparent diffusion | Bimolecular recombination |
|---|---|---|---|---|
| $D_1\ (cm^2 s^{-1})$ | Various (0.05, 0.5, 1, or time-varying (Eq. 21)) | 1 | 0.5 | 0 or 0.1 |
| $k_{d1}\ (ns^{-1})$ | 1 | 1 | 2 | 1 |
| $\phi_1$ | 1 | 1 | 2 | 1 |
| $N_{1,init}$ | 1 | 0.5 | 1 | 1 |
| $D_2\ (cm^2 s^{-1})$ | --- | 0 | 0 | --- |
| $k_{d2}\ (ns^{-1})$ | --- | 1 | 0.1 | --- |
| $\phi_2$ | --- | 1 | 0.1 | --- |
| $N_{2,init}$ | --- | 0.5 | 0 | --- |
| $k_{12}\ (ns^{-1})$ | --- | 0 | 2 | --- |
| $k_{21}\ (ns^{-1})$ | --- | 0 | 0 | --- |
| $k_{bi}\ (nm^3 ns^{-1})$ | --- | --- | --- | 1 |

*Table S2. Physical parameters used for finite difference simulations.*

| **Parameter** | |
|---|---|
| Simulation radius (nm) | 15000 |
| Simulation $\Delta x$ (nm) | 5 |
| Simulation end time (ns) | 3 (single species diffusion, no noise)<br>10 (single species diffusion, noise added)<br>3.5 (two species diffusion)<br>5 (bimolecular recombination) |
| Simulation $\Delta t$ (ps) | 0.1 |
| Data collection radius (nm) | 2500 |
| Data collection $\Delta x$ (nm) | 25 |
| Data collection $\Delta t$ (ps) | 10 |
| Data collection end time (ns) | *Same as simulation end time* |

*Table S3. Parameters used for simulating numerical solutions using finite differences and dataset generation.*

Where appropriate, random noise was added to the simulated datasets according to the following procedure. First, the measured counts were scaled such that the signal at the center of the profile at $t = 0$ was set to 900 counts in the low-noise case, and 100 counts in the high-noise case. Then, to simulate shot noise, the signal in each space and time bin was determined by sampling a Poisson distribution with the mean set to the corresponding noise-free value. Since the standard deviation of a Poisson random variable is the square root of the mean, this meant that the peak signal-to-noise ratio was 30 and 10 in the low- and high-noise cases, respectively. Finally, a position-independent term was added by sampling a white-noise term to each space and time bin, each sampled from a Poisson distribution with a mean and variance of 1.

Once the simulated datasets were finalized, the MSDs at each time point were calculated using the methods outlined in the main text. For noisy data, the background was estimated using the data at later times and subtracted from profiles before calculating MSDs. For the curve-fitting methods, the normalized profiles were fit using non-linear least squares optimization. The diffusivities for each method were then determined by using linear regression to fit a line to the MSD evolution over the first nanosecond, then applying the relationship $MSD = 2Dt$.

### S1.3. KMC simulations

Ballistic transport and photon recycling were simulated as continuous-time random walks using kinetic Monte Carlo (KMC) techniques. The walkers represented charge carriers in the ballistic transport case and excitons in the photon recycling case. The *x*-positions of the walkers were initialized by randomly sampling from the initial profile distributions outlined in section S1.1. Then, for each walker, a step time was randomly drawn from an exponential distribution with a mean equal to the scattering time of 500 fs (for ballistic transport) or the radiative lifetime of 1 ns (for photon recycling). For ballistic transport, the step distance was calculated from the step time using the carrier speed of 300 km $s^{-1}$. For photon recycling, the step distance was calculated by sampling a random wavelength from the PL distribution, then sampling a random exponential variable with a mean set to the absorption distance at that wavelength (see Fig. S4). The *x*- component of the step was calculated assuming the step occurred along a random direction in three dimensions, and the new $x$-position and time for the walker were recorded. This process was repeated for each walker until the total time of its steps surpassed the simulation time, or until the walker left the simulation radius, at which point it was discarded.

Once the KMC simulation was finished, the positions of the walkers were sampled at regular time intervals and binned spatially to generate the final datasets. For the ballistic transport simulations, the position at a given time was calculated by linearly interpolating between the positions before and after the sample time. For the photon recycling simulations, the position at a given time was set to the position of the last step before the sample time, as the photon emission and reabsorption process is effectively instantaneous. The parameters for the ballistic transport and photon recycling simulations are given in table S4.

| **Parameter** | Ballistic Transport | Photon Recycling |
|---|---|---|
| # of walkers | $5 \times 10^6$ | $5 \times 10^6$ |
| Simulation radius (nm) | 6000 | 6000 |
| Simulation end time | 3.1 ps | 3.1 ns |
| Data collection radius (nm) | 2500 | 5000 |
| Data collection $\Delta x$ (nm) | 25 | 25 |
| Data collection $\Delta t$ | 10 fs | 10 ps |
| Data collection end time | *Same as simulation end time* | *Same as simulation end time* |

*Table S4. Simulation and data collection parameters used for kinetic Monte Carlo calculations.*

## S2. Analytical solution to two-species diffusion with interchange

As discussed in the main text, we model the evolution of a system with two diffusing species $n_1$ and $n_2$, each having independent decay and interchange terms, using the following system of partial differential equations:

$$\frac{\partial n_1(x,t)}{\partial t} = D_1 \frac{\partial^2 n_1(x,t)}{\partial x^2} - (k_{d1} + k_{12})\, n_1(x,\, t) + k_{21}\, n_2(x,\, t) \tag{S3a}$$

$$\frac{\partial n_2(x,t)}{\partial t} = D_2 \frac{\partial^2 n_2(x,t)}{\partial x^2} - (k_{d2} + k_{21})\, n_2(x,\ t) + k_{12}\, n_1(x,\ t) \tag{S3b}$$

$$n_1(x,\ 0) = N_{1,init}\, \delta(x), n_2(x,\ 0) = N_{2,init} \delta(x) \tag{S3c,d}$$

And we model the signal from this system $q(x,t)$ using the following equation:

$$q(x,\ t) = \Phi_1 n_1(x,t) + \Phi_2 n_2(x,t) \tag{S4}$$

The solution to this system of equations in reciprocal space was calculated using the Symbolic Math Toolbox in MATLAB™. We first define the following quantities:

$$\lambda_1 = ((D_1 - D_2)^2 k^4 + 2(D_1 - D_2)(k_{d1} - k_{d2} + k_{12} - k_{21})k^2 + (k_{d1} - k_{d2})^2 + 2(k_{d1} - k_{d2})(k_{12} - k_{21}) + (k_{12} + k_{21})^2)^{\frac{1}{2}} \tag{S5a}$$

$$\lambda_2 = \exp\left[-\frac{t}{2}(k_{d1} + k_{d2} + k_{12} + k_{21} + (D_1 + D_2)k^2)\right] \tag{S5b}$$

$$\lambda_3 = \sinh\left[\frac{\lambda_1}{2} t\right] \tag{S5c}$$

$$\lambda_4 = \cosh\left[\frac{\lambda_1}{2} t\right] \tag{S5d}$$

$$\lambda_5 = k_{d2} - k_{d1} - k_{12} + k_{21} + (D_2 - D_1)k^2 \tag{S5e}$$

The recorded signal is then:

$$\hat{q}(k,\ t) = N_{1,init}\Phi_1\lambda_2\left(\lambda_4 + \frac{\lambda_3}{\lambda_1}\left(\lambda_5 + 2k_{21}\frac{N_{2,init}}{N_{1,init}}\right)\right) + N_{2,init}\Phi_2\lambda_2\left(\lambda_4 + \frac{\lambda_3}{\lambda_1}\left(-\lambda_5 + 2k_{12}\frac{N_{1,init}}{N_{2,init}}\right)\right) \tag{S6}$$

Since the system outlined in Eq. S3a-b is linear, the dynamics for any initial distribution in real space are given by convolving the original profile shape with the impulse-response solution (Eq. S6). Therefore, by the convolution theorem, multiplying any initial profile in reciprocal space by Eq. S6 will give its evolution in reciprocal space.

## S3. Quantitative analysis of ballistic transport

Say a population of excited-state species undergoes ballistic transport and has an initial probability distribution of $n(x,0) = \delta(x)$, and say we collect a dataset by recording the spatiotemporal distribution of the excited-state species along the $x$-direction, as in Fig. 9a-d. Let $p_v(v_x)$ be the probability distribution function for the $x$-component of its velocity distribution $v_x$. We assume that $p_v(v_x)$ is symmetric around the origin. At times much shorter than the characteristic scattering time, the spatial distribution can be approximated by assuming that no scattering events have occurred. Under this assumption, the spatial probability distribution can be directly related to the velocity distribution according to:

$$n(x,t) = \frac{1}{t} p_v\left(\frac{x}{t}\right) \tag{S7}$$

The Fourier transform of Eq. S7 is given by:

$$\mathcal{F}\{n(x,t)\} = \hat{n}(k,t) = \int_{-\infty}^{\infty} \frac{1}{t} p_v\left(\frac{x}{t}\right) \exp[-ikx]dx \tag{S8}$$

Under the assumptions we made about the initial profile shape, $\hat{n}(k,0) = 1$. Also, the total population is constant at 1, $\hat{n}(0,t) = 1$. Therefore, we can directly equate Eq. S8 to the normalized spatial mode decays $f(k,t)$ used elsewhere in the text.

$$\hat{n}(k,t) = f(k,t) = \int_{-\infty}^{\infty} \frac{1}{t} p_v\left(\frac{x}{t}\right) \exp[-ikx]dx \tag{S9}$$

To estimate the behavior of $f(k,t)$ at small values of $k$, we evaluate the Taylor expansion around $k = 0$:

$$f(k,t) \approx f(0,t) + k\left(\frac{\partial}{\partial k} f(k,t)\right)\Bigg|_{k=0} + \frac{k^2}{2}\left(\frac{\partial^2}{\partial k^2} f(k,t)\right)\Bigg|_{k=0} + O(k^3) \tag{S10}$$

$f(0,t) = 1$ by construction. Next, we evaluate the linear term by substituting in Eq. S9:

$$\begin{aligned}\left(\frac{\partial}{\partial k} f(k,t)\right)\Bigg|_{k=0} &= \left(\frac{\partial}{\partial k}\int_{-\infty}^{\infty}\frac{1}{t}p_v\left(\frac{x}{t}\right)\exp[-ikx]\,dx\right)\Bigg|_{k=0} \\ &= \left(\int_{-\infty}^{\infty}\frac{1}{t}p_v\left(\frac{x}{t}\right)\frac{\partial}{\partial k}\exp[-ikx]\,dx\right)\Bigg|_{k=0} \\ &= \left(\int_{-\infty}^{\infty}\frac{1}{t}p_v\left(\frac{x}{t}\right)(-ix)\exp[-ikx]\,dx\right)\Bigg|_{k=0}\end{aligned} \tag{S11}$$

Evaluating this term at $k = 0$ yields:

$$\left(\frac{\partial}{\partial k} f(k,t)\right)\Bigg|_{k=0} = -i\int_{-\infty}^{\infty}\frac{1}{t}p_v\left(\frac{x}{t}\right)x\,dx \tag{S12}$$

By comparison with Eq. S7, we can see that Eq. S12 is proportional to the mean of the population distribution at any given time. However, since our population evolves symmetrically, its mean, and therefore the linear term in the Taylor expansion, must be 0. This implies that the initial slope for each normalized spatial mode is 0.

By manipulating the quadratic term using the same process illustrated in Eq. S11-12, we can show that:

$$\begin{aligned}\left(\frac{\partial^2}{\partial k^2} f(k,t)\right)\Bigg|_{k=0} &= \left(\int_{-\infty}^{\infty}\frac{1}{t}p_v\left(\frac{x}{t}\right)(-ix)^2\exp[-ikx]\,dx\right)\Bigg|_{k=0} \\ &= -\int_{-\infty}^{\infty}\frac{1}{t}p_v\left(\frac{x}{t}\right)x^2\,dx\end{aligned} \tag{S13}$$

By comparison with Eq. S7, we can see that Eq. S13 is the negative variance of the spatial distribution $n(x,t)$ at any given time, which is equivalent to the negative MSD since we are considering the impulse response of the system. Substituting in $v_x = x/t$ yields:

$$\begin{aligned}\left(\frac{\partial^2}{\partial k^2} f(k,t)\right)\Bigg|_{k=0} &= -MSD = -\int_{-\infty}^{\infty}\frac{1}{t}p_v(v_x)\,(v_x t)^2\,t\,dv_x \\ &= -t^2\int_{-\infty}^{\infty}p_v(v_x)\,v_x^2\,dv_x\end{aligned} \tag{S14}$$

Since $p_v(v_x)$ is symmetric around the origin, we can use the definition of variance to write:

$$-t^2 \int_{-\infty}^{\infty} p_v(v_x)\, v_x^2 \, dv_x = -t^2 Var[v_x] \quad \text{(S15)}$$

Comparing Eq. S14 and Eq. S15, we can write that, at early times:

$$MSD = t^2 Var[v_x] \quad \text{(S16)}$$

And substituting Eq. 15 into the Taylor expansion in Eq. S10 yields:

$$f(k,t) \approx 1 - \frac{1}{2} k^2 t^2 Var[v_x] \quad \text{(S17)}$$

To illustrate this relationship described in Eq. S17, we fit the first 100 *fs* of our simulated normalized spatial mode decay curves from Fig. 9c-d to the expression $1 - \Lambda t^2$. As shown in Fig. S3, the resulting quadratic decay rates followed the relationship $\Lambda = \frac{1}{2} Var[v_x] k^2$ predicted by Eq. S17.

## S4. Decay rate scaling for Lévy flights in reciprocal space

Given that the duration of our photon recycling simulations is similar to our simulated radiative lifetime of 1 *ns*, it may be surprising that the normalized spatial modes for this system (Fig. 9g) resemble those of a Lévy-stable density function throughout the entire temporal range. However, this relation follows from the exponential function used to model the photon recycling rate. For a continuous-time random walk with a step time distribution $\psi(t)$ and step size distribution $p(x)$, the distribution of the walkers can be expressed in reciprocal time and space as:[1]

$$\hat{\tilde{f}}(k,\, s) = \frac{1 - \tilde{\psi}(s)}{s\left(1 - \tilde{\psi}(s)\hat{p}(k)\right)} \quad \text{(S18)}$$

If we assume an exponential step time distribution $\psi(t) = k_{rad} \exp[-k_{rad} t]$, then one can perform an inverse Laplace transform on Eq. S18 to show that:

$$\hat{f}(k,\, t) = \exp\left[-k_{rad}\left(1 - \hat{p}(k)\right)t\right] \quad \text{(S19)}$$

Eq. S19 shows that the normalized spatial modes for our photon recycling simulation should follow exponential decays, as we see in Fig. 9g. In this case, the spatial mode decay rates can also be directly related to the step size distribution through:

$$\Gamma(k) = k_{rad}\big(1 - \hat{p}(k)\big) \tag{S20}$$

If $p(x)$ has power-law tails that are proportional to $|x|^{-1-\alpha}$ in the large-*x* limit, with $0 < \alpha < 2$, then in the corresponding small-*k* limit:[1]

$$\hat{p}(k) = \exp[-b|k|^{\alpha}] \tag{S21}$$

Here, $b$ is a constant. Substituting Eq. S21 into Eq. S20, and expanding around $k = 0$ gives the following expression for $\Gamma(k)$ in the small-*k* limit:

$$\Gamma(k) \approx k_{rad} b|k|^{\alpha} = \lambda|k|^{\alpha} \tag{S22}$$

Eq. S22 is the same as Eq. 24 in the main text. However, Eq. S22 was derived from the step size distribution, without assuming convergence to a Lévy-stable distribution. The "small-*k*" limit where Eq. S22 applies depends on the distances where $p(x)$ converges to power-law behavior. Given that photon recycling is a long-range transport mechanism, this may not happen in the range of a dataset.

## S5. Early-time behavior of 2 diffusing species in reciprocal space

The normalized spatial modes for the system of two diffusing species outlined in section S2 are given by $\hat{q}(k,\ t)/\hat{q}(0,t)$ . At early times, the change in the normalized spatial modes can be modeled by a first-order Taylor expansion around $t = 0$. In the case of two diffusing species, this Taylor expansion can be expressed as:

$$\frac{\hat{q}(k,\ t)}{\hat{q}(0,t)} \approx 1 - \frac{N_{1,init}\Phi_1 D_1 + N_{2,init}\Phi_2 D_2}{N_{1,init}\Phi_1 + N_{2,init}\Phi_2}\ k^2 t \tag{S23}$$

Eq. S23 demonstrates that the early-time evolution of normalized spatial modes for a system with two diffusing species is equivalent to that of a single diffusing species with an effective diffusivity $D_{eff}$ of:

$$D_{eff} = \frac{N_{1,init}\Phi_1 D_1 + N_{2,init}\Phi_2 D_2}{N_{1,init}\Phi_1 + N_{2,init}\Phi_2} \tag{S24}$$

Therefore, at early times, the slopes of the normalized spatial modes for two diffusing species preserve the quadratic relationship with $k$ expected for a single species.

## S6. Supplementary Figures

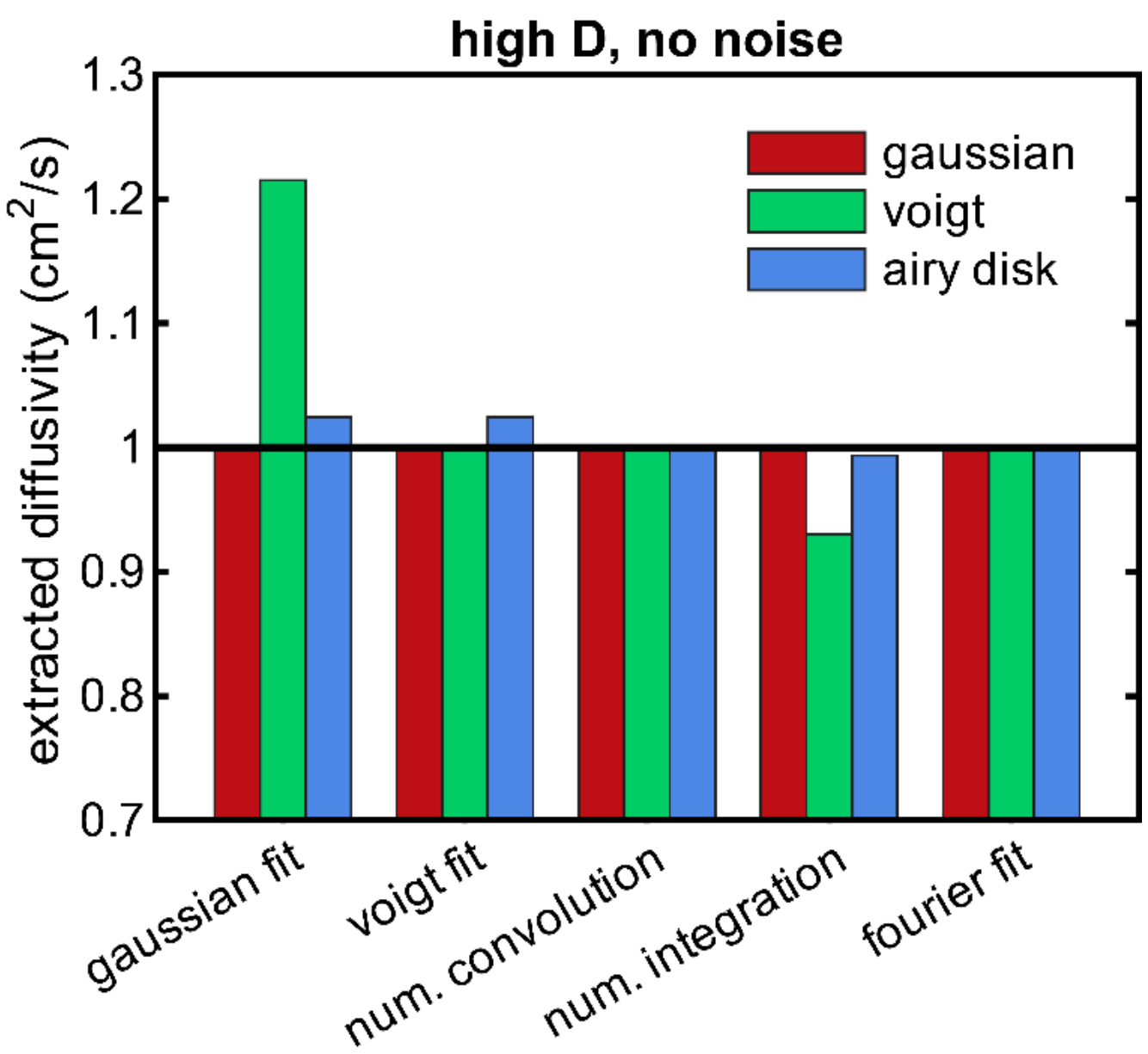

**Figure S1.** Extracted diffusivity values for each MSD calculation method (Gaussian fit, Voigt fit, numerical convolution, numerical integration, and Fourier fit) and initial profile shape (Gaussian, Voigt, and Airy disk) in Cartesian geometry for a simulated diffusivity value of 1 cm$^2$/s (horizontal line) and with no added noise.

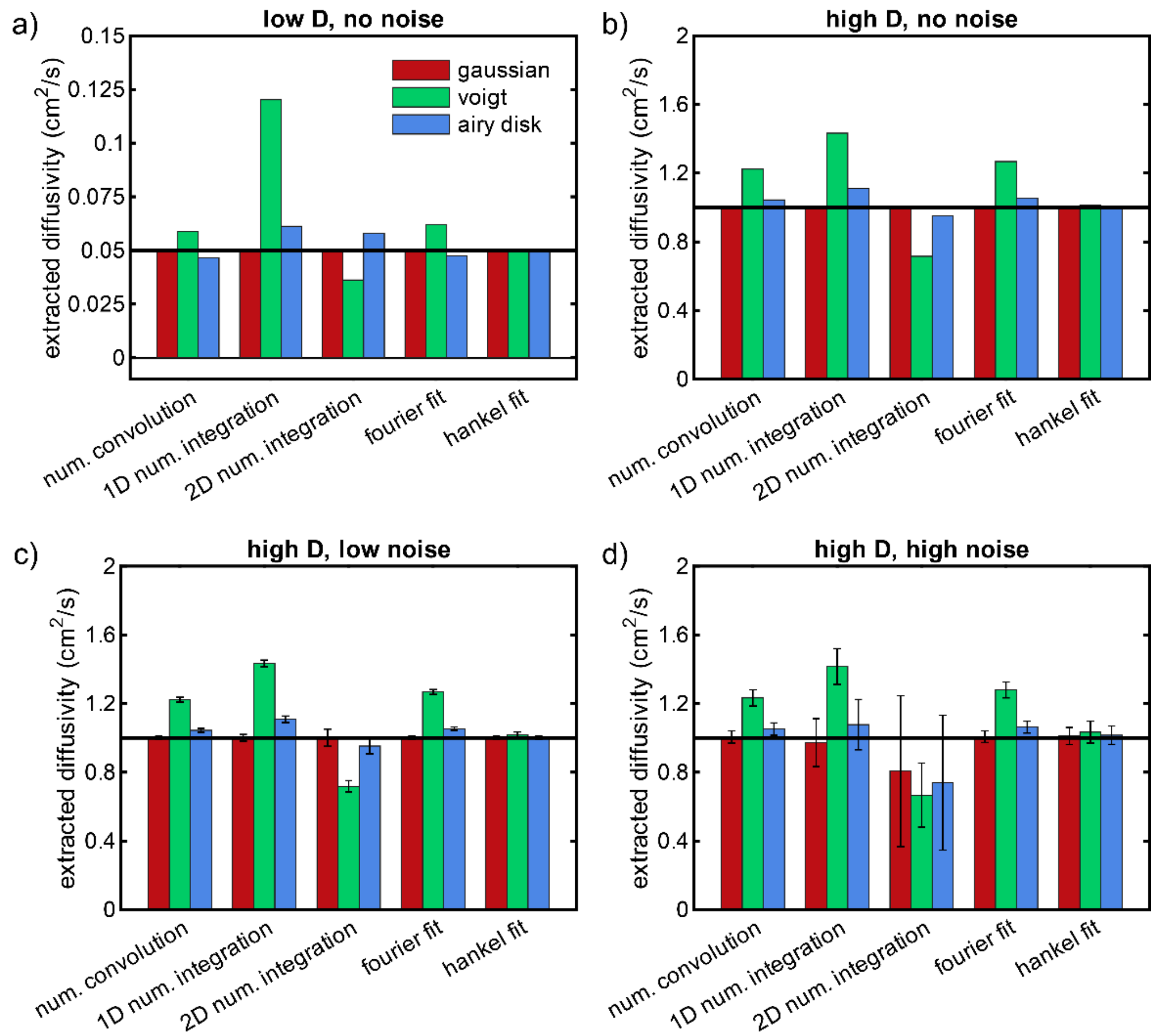

**Figure S2.** Extracted diffusivity values for each MSD calculation method (Gaussian fit, Voigt fit, numerical convolution, numerical integration, and Fourier fit) and initial profile shape (Gaussian, Voigt, and Airy disk) in cylindrical geometry. Bar heights represent the average extracted diffusivity value, and the error bars, when present, denote the standard deviation. The horizontal lines at 0.05 $cm^2/s$ (a) or 1 $cm^2/s$ (b-d) denote the true diffusivity value used in the simulation. The colors for each initial profile shape are consistent for all bar charts. Each panel represents a different combination of diffusivity value and noise level: a) low-diffusivity, no-noise; b) high-diffusivity, no noise; c) high-diffusivity, low-noise; d) high-diffusivity, high-noise.

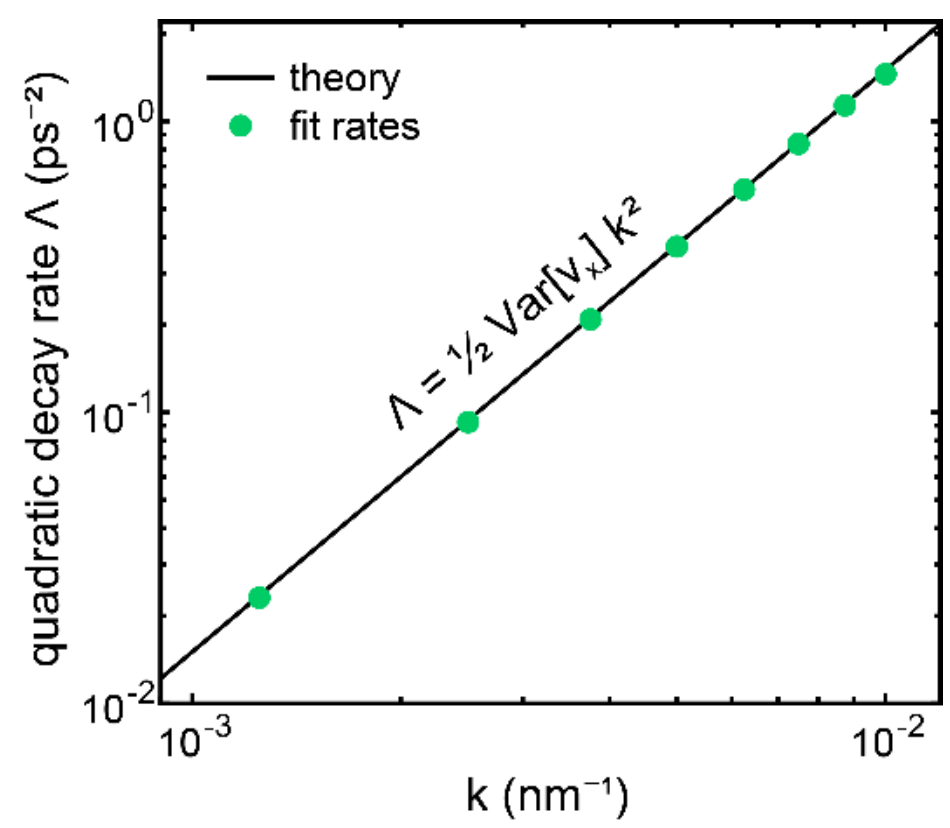

**Figure S3.** Fits to the initial quadratic decay rates in the first 100 *fs* as a function of *k* for the normalized spatial modes plotted in Fig. 9c-d. The solid black line depicts their theoretical relationship for the simulated velocity profile.

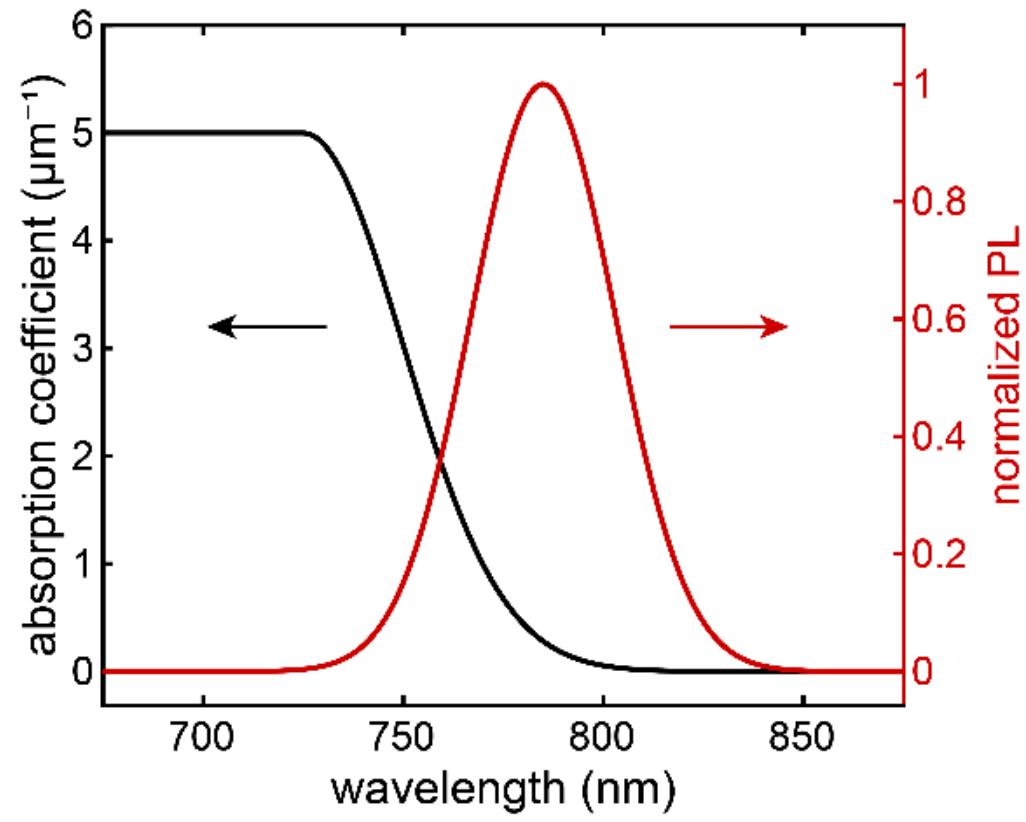

**Figure S4.** Absorption and PL spectra used for the KMC simulation of photon recycling.

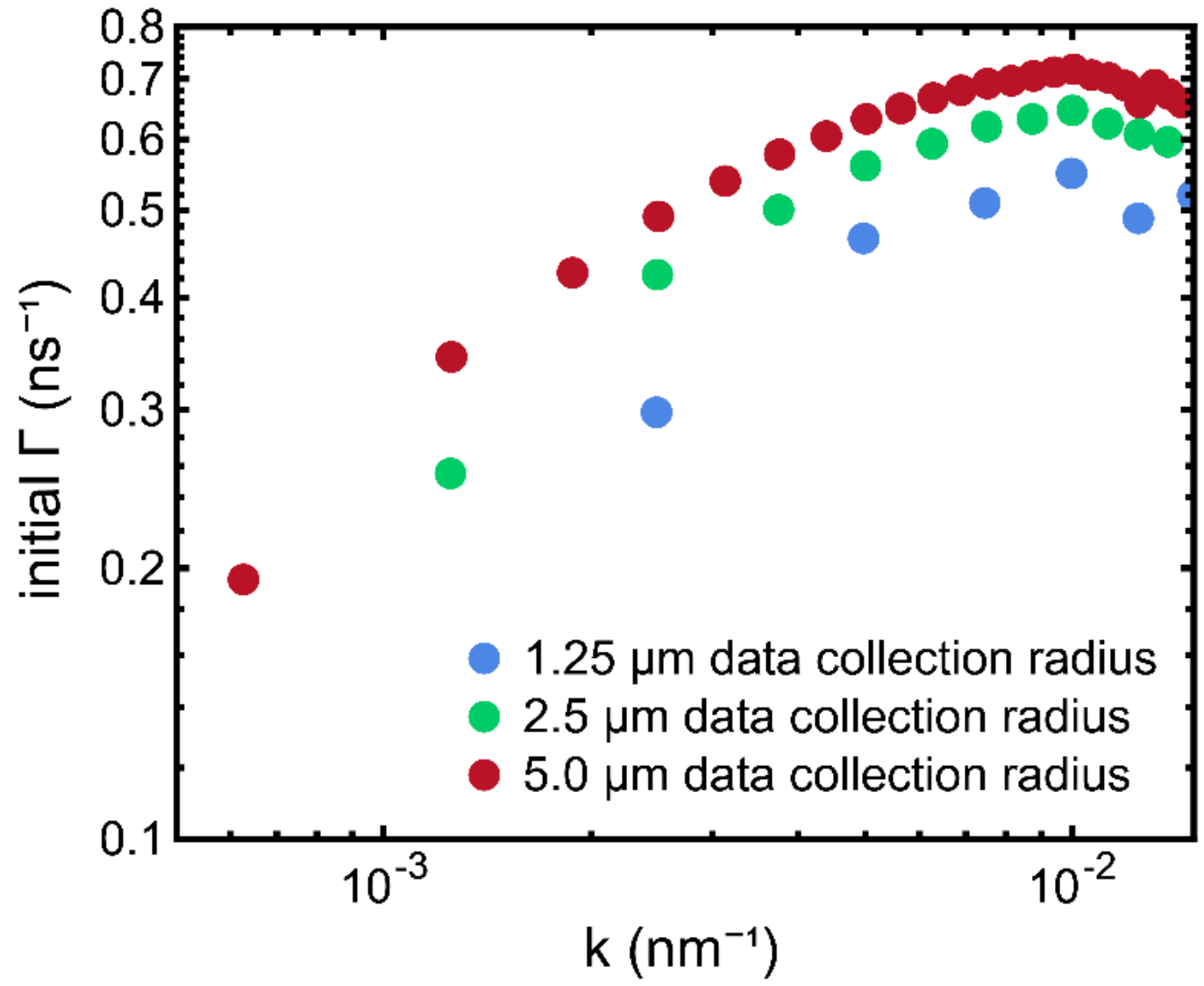

**Figure S5.** Initial decay rates of spatial modes within the first 50 ps as a function of *k* for photon recycling simulations with data collection radii of 1.25 µm, 2.5 µm, and 5.0 µm. Due to the long-distance nature of photon recycling transport, the extracted spatial mode decay rates vary as the width of the data collection region changes, especially at the lowest value of *k* for each collection radius.

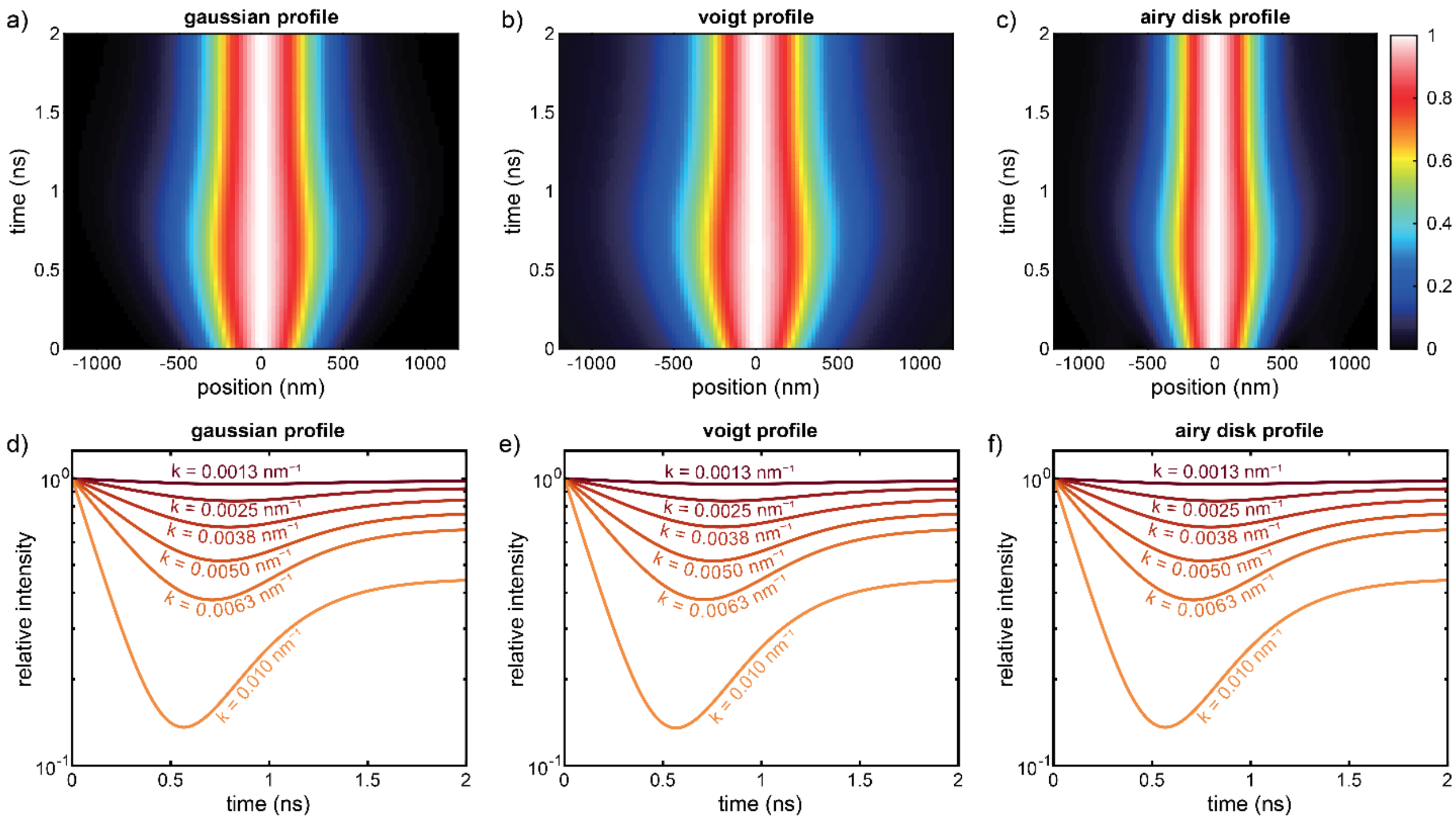

**Figure S6.** *Two-species diffusion with different initial profiles.* a-c) Normalized color map depicting the spatiotemporal evolution of a system of two diffusing species exhibiting negative effective diffusivity with a Gaussian, Voigt, and Airy disk initial profile, respectively. All simulations are identical except for the initial profile shape. d-f) Normalized spatial mode decays for the simulations displayed in panels a-c with d) Gaussian, e) Voigt, and f) Airy disk initial profile shapes. The evolution of the normalized spatial modes is almost identical for each initial profile.

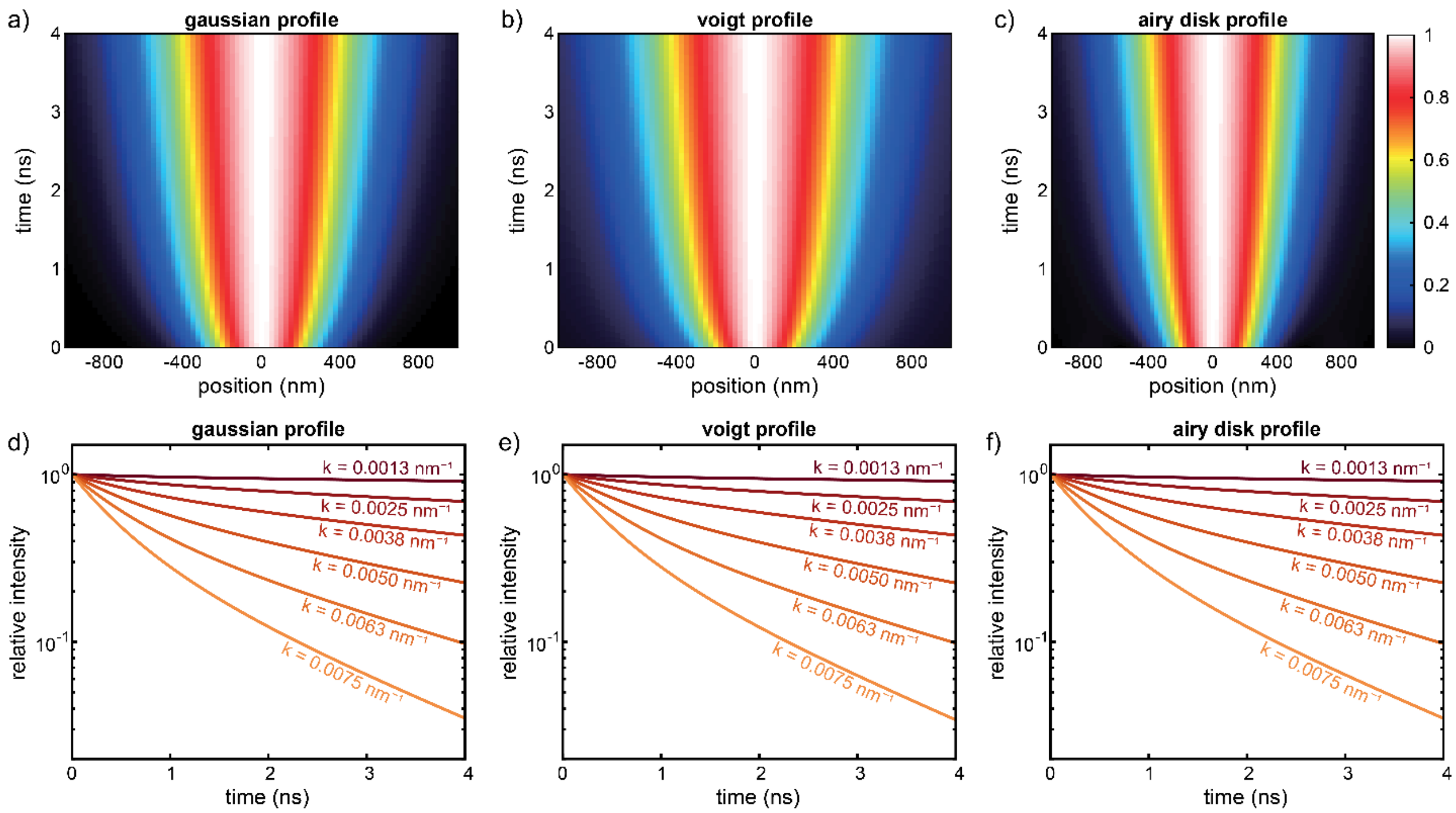

**Figure S7.** *Out-of-equilibrium diffusion appearing subdiffusive with different initial profiles.* a-c) Normalized color map depicting the spatiotemporal evolution of a system having a diffusivity that decays in time with a Gaussian, Voigt, and Airy disk initial profile, respectively. All simulations are identical except for the initial profile shape. d-f) Normalized spatial mode decays for the simulations displayed in panels a-c with d) Gaussian, e) Voigt, and f) Airy disk initial profile shapes. The evolution of the normalized spatial modes is almost identical for each initial profile.

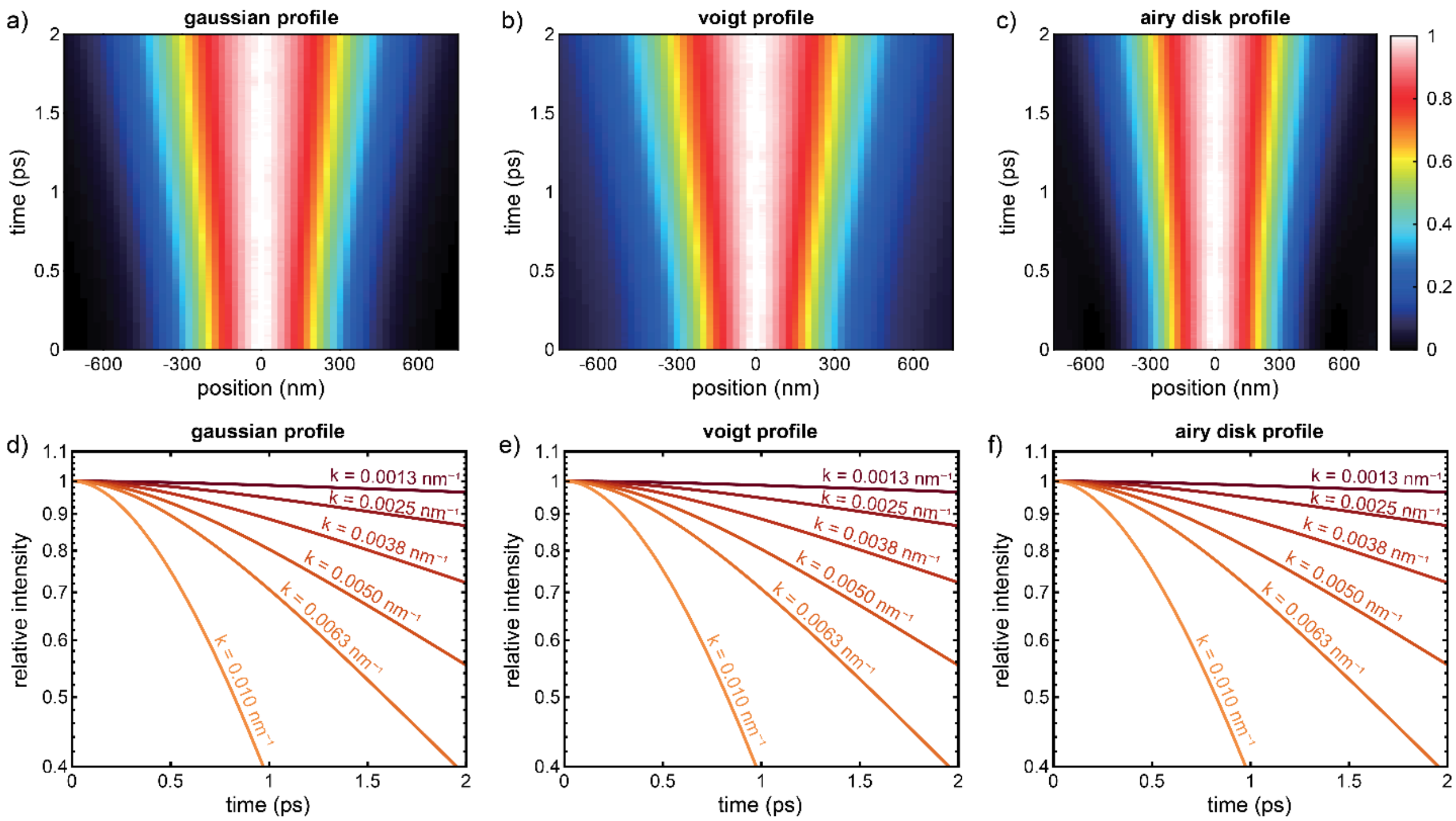

**Figure S8.** *Ballistic transport with different initial profiles.* a-c) Normalized color map depicting the spatiotemporal evolution of a system undergoing ballistic transport with a Gaussian, Voigt, and Airy disk initial profile, respectively. All simulations are identical except for the initial profile shape. d-f) Normalized spatial mode decays for the simulations displayed in panels a-c with d) Gaussian, e) Voigt, and f) Airy disk initial profile shapes. The evolution of the normalized spatial modes is almost identical for each initial profile.

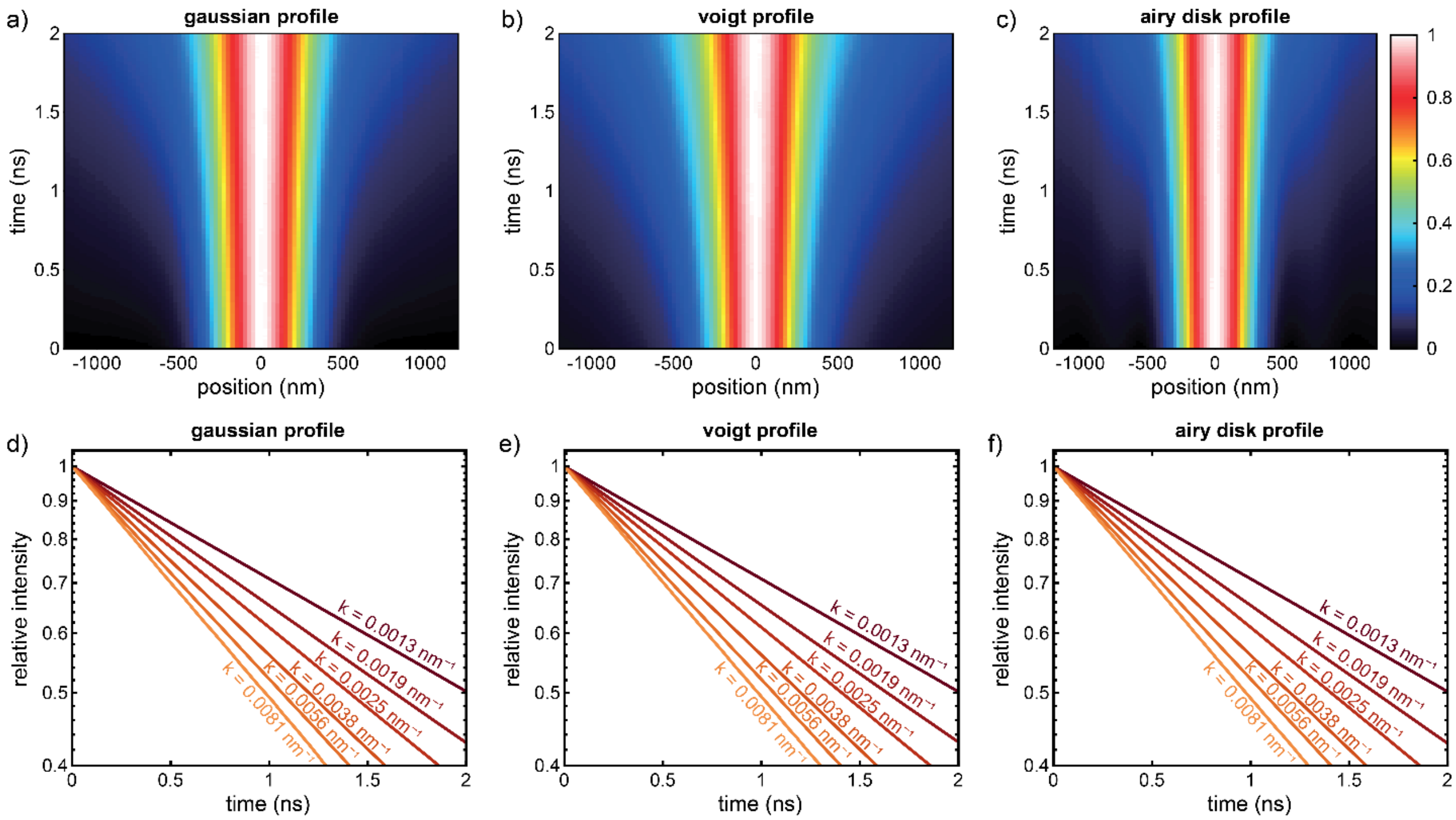

**Figure S9.** *Photon recycling with different initial profiles.* a-c) Normalized color map depicting the spatiotemporal evolution of a system undergoing photon recycling with a Gaussian, Voigt, and Airy disk initial profile, respectively. All simulations are identical except for the initial profile shape. d-f) Normalized spatial mode decays for the simulations displayed in panels a-c with d) Gaussian, e) Voigt, and f) Airy disk initial profile shapes. The evolution of the normalized spatial modes is almost identical for each initial profile.

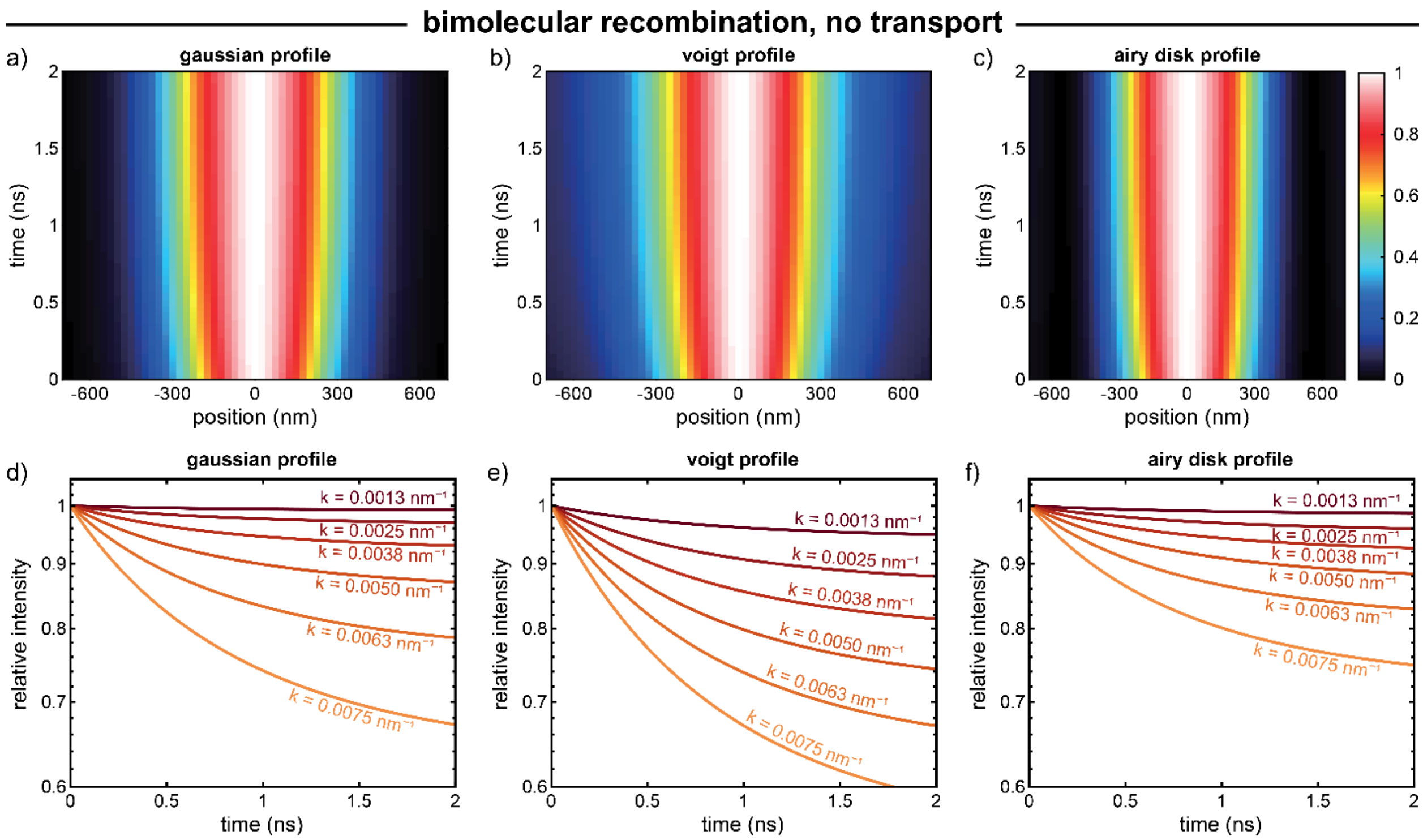

**Figure S10.** *Bimolecular recombination of a stationary species with different initial profiles.* a-c) Normalized color map depicting the spatiotemporal evolution of a system undergoing bimolecular recombination and no diffusion with a Gaussian, Voigt, and Airy disk initial profile, respectively. All simulations are identical except for the initial profile shape. d-f) Normalized spatial mode decays for the simulations displayed in panels a-c with d) Gaussian, e) Voigt, and f) Airy disk initial profile shapes. Unlike earlier simulations, the evolution of the normalized spatial modes is noticeably different for each initial profile.

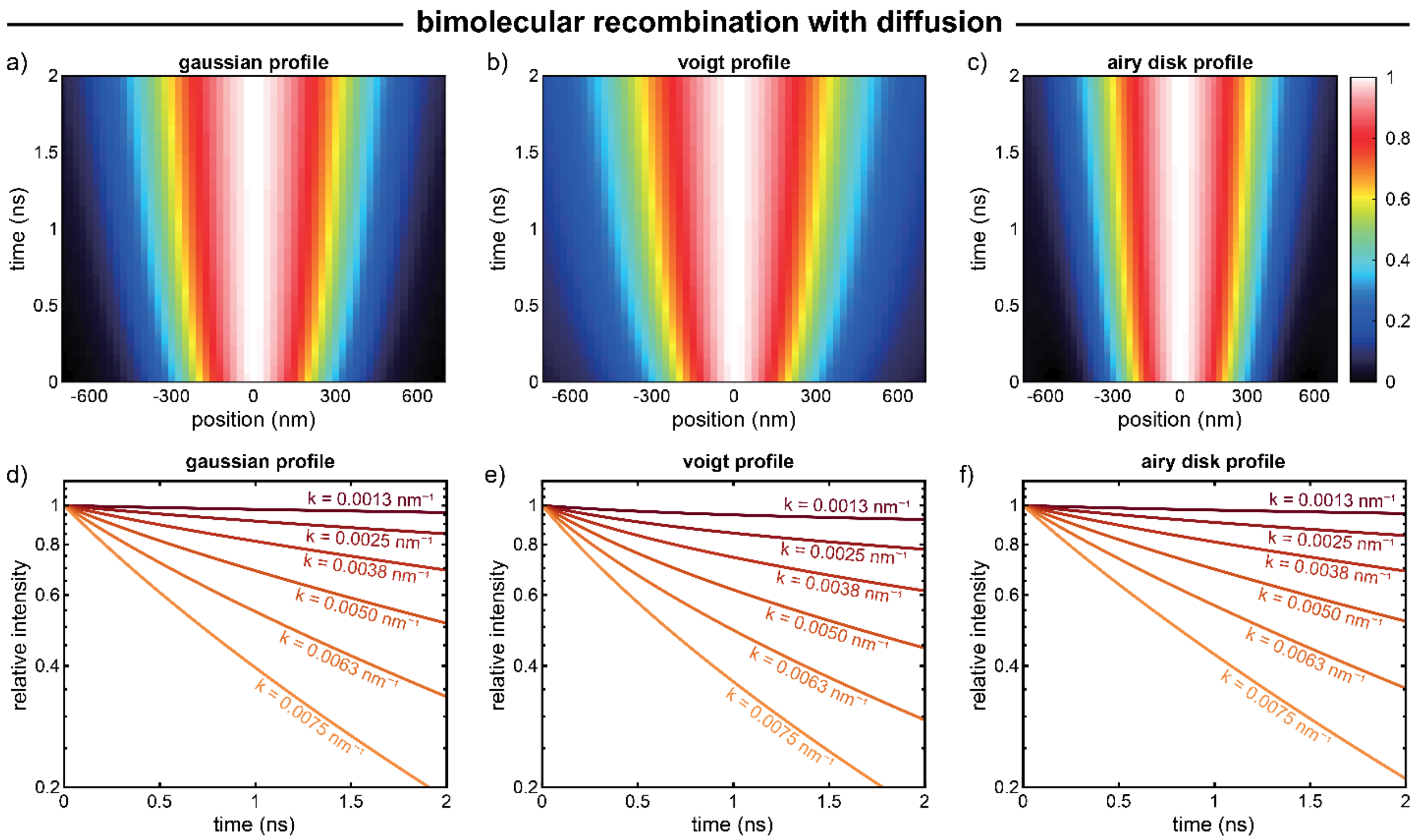

**Figure S11.** *Bimolecular recombination of a species undergoing diffusion with different initial profiles.* a-c) Normalized color map depicting the spatiotemporal evolution of a system undergoing bimolecular recombination and diffusion with $D = 0.1\ cm^2/s$ and having a Gaussian, Voigt, and Airy disk initial profile, respectively. All simulations are identical except for the initial profile shape. d-f) Normalized spatial mode decays for the simulations displayed in panels a-c with d) Gaussian, e) Voigt, and f) Airy disk initial profile shapes. The evolution of the normalized spatial modes is noticeably different for each initial profile.